\documentclass[12pt]{article}

\usepackage{cite}
\usepackage{mathptmx}
\usepackage{setspace}
\usepackage{multirow}
\usepackage{longtable}

\usepackage{changepage}
\usepackage{dcolumn}
\usepackage[table]{xcolor}
\usepackage{floatrow}
\usepackage{float}
\usepackage{subcaption}
\usepackage{graphicx,xcolor} 
\usepackage{url}
\usepackage{hyperref}

\usepackage[margin=1in]{geometry}
\usepackage{amsmath}
\usepackage{amssymb}
\usepackage{pgf}

\definecolor{DarkCyan}{rgb}{ 0 0.545 0.937}
\definecolor{LightCyan}{rgb}{ 0.28 0.82 0.8}
\definecolor{myblue-strong}{rgb}{0.4 0.67 0.84}
\definecolor{myblue-medium}{HTML}{A3CCE7}
\definecolor{myblue-weak}{HTML}{D1E5F3}
\definecolor{myblue-negligible}{HTML}{EFF6FB}

\definecolor{table-color-dark}{HTML}{E4D2BA}
\definecolor{table-color-light}{HTML}{F6F0E8}

\definecolor{myred-strong}{rgb}{0.91 0.60 0.46}
\definecolor{myred-medium}{HTML}{F1C0AB}
\definecolor{myred-weak}{HTML}{FAEAE3}
\definecolor{myred-negligible}{HTML}{FCF4F1}

\newcolumntype{d}[1]{D..{#1}}

\DeclareMathOperator{\EX}{\mathbb{E}}

\title{Correlates of the country differences in the infection and mortality rates during the first wave of the COVID-19 pandemic: Evidence from Bayesian model averaging
}
\author{Viktor Stojkoski$^{1,2,\footnote{Corresponding author: \href{mailto:vstojkoski@eccf.ukim.edu.mk}{vstojkoski@eccf.ukim.edu.mk} }}$, Zoran Utkovski$^{3,2}$,  Petar Jolakoski$^{1}$, \\ Dragan Tevdovski$^{1}$, and Ljupco Kocarev$^{2,4}$ \\ \\  $^{1}$Faculty of Economics, Ss. Cyril and Methodius University in Skopje \\
$^{2}$Macedonian Academy of Sciences and Arts \\ 
$^{3}$Fraunhofer Heinrich Hertz Institute, Berlin \\
$^{4}$Faculty of Computer Science and Engineering, \\ Ss. Cyril and Methodius University in Skopje}
\date{\today}

\begin{document}
\maketitle

\begin{abstract}
In the initial wave of the COVID-19 pandemic we observed great discrepancies in both infection and mortality rates between countries. Besides the biological and epidemiological factors, a multitude of social and economic criteria also influence the extent to which these discrepancies appear. Consequently, there is an active debate regarding the critical socio-economic and health factors that correlate with the infection and mortality rates outcome of the pandemic. Here, we leverage Bayesian model averaging techniques and country level data to investigate the potential of 28 variables, describing a diverse set of health and socio-economic characteristics, in being correlates of the final number of infections and deaths during the first wave of the coronavirus pandemic. We show that only few variables are able to robustly correlate with these outcomes. To understand the relationship between the potential correlates in explaining the infection and death rates, we create a Jointness Space. Using this space, we conclude that the extent to which each variable is able to provide a credible explanation for the COVID-19 infections/mortality outcome varies between countries because of their heterogeneous features.
\end{abstract}

\section{Introduction}

In order to reduce the potential enormous impact of the coronavirus disease spread (COVID-19), most governments implemented social distancing restrictions such as closure of schools, airports, borders, restaurants and shopping malls. In the most severe cases there were even lockdowns -- all citizens were prohibited from leaving their homes. This subsequently led to a major economic downturn: stock markets plummeted, international trade slowed down, businesses went bankrupt and people were left unemployed. While in some countries the implemented restrictions had a significant impact on reducing the expected shock from the coronavirus, the extent of the disease spread in the population greatly varied from one economy to another.

A multitude of health, social and economic factors have been attributed as potential correlates for the observed variety in the coronavirus outcome in terms of the number of infections and/or deaths during this first wave of the pandemic. Indeed, there are numerous studies which discover various factors that affect the \textit{within} country distribution of infections and deaths (See for example, Refs.~\cite{singu2020impact,galanis2021incorporating,rollston2020covid,kleitman2021comply,clouston2021socioeconomic}). The same debate has been extended to evaluate the \textit{between} country discrepancies. In particular, some experts say that the hardest hit countries also had an aging population~\cite{gardner2020coronavirus,lima2020emotional}, or an underdeveloped healthcare system~\cite{tanne2020covid,mikhael2020can}. Others emphasize the role of the natural environment~\cite{di2020opinion,wu2020exposure}. Having in mind the ongoing discussion, a comprehensive empirical study of the critical health, social and economic correlates with the country level outcome of the number of infections and deaths during first wave of the pandemic can not only aid in inferring whether there are any general rules in their potential impact, but would also offer a guidance for future policies that aim at preventing the emergence of future epidemic crises.

To this end, here we perform a detailed statistical analysis on a large set of potential health and socio-economic variables and explore their potential to explain the variety in the observed coronavirus total infections/deaths between countries in the \textit{first wave} of the virus spread.We focus on COVID-19 data that is generated only in the first wave of the pandemic, and thus do not account for various waves (we formally define the first wave in the next section). While this may be seen as a limitation of our analysis, we assert that for each subsequent wave, there was larger knowledge for the spread of the virus and vaccines were available. This significantly impacted in the way in which the population reacted to the potential susceptibility. Thus, it can be said that each wave has its own health, social and economic characteristics and therefore it should be studied separately.

To construct the set of potential correlates we conduct a thorough review of the literature that describes the social and economic factors which contribute to the spread of an epidemic.We identify a total of 28 potential variables that describe a diverse ensemble of factors, including: healthcare infrastructure, societal characteristics, economic performance, demographic structure etc. To investigate the performance of each variable in explaining the coronavirus infections/deaths outcome, we collect a sample of 105 countries, the largest set of countries for which all data were available, and utilize the technique of Bayesian model averaging (BMA). BMA allows us to isolate the most important correlates by calculating the posterior probability that they truly regulate the process. At the same time, BMA provides estimates for the relative impact of the correlates and accounts for the uncertainty in their selection~\cite{raftery1997bayesian,hoeting1999bayesian,sala2004determinants}.  In this aspect, our analysis adds value a growing body of literature which applies Bayesian methods for investigating the critical factors that drive a certain process, and in this particular case the outcome of the COVID-19 pandemic~\cite{fragoso2018bayesian}.

Based on the studied data, we observe patterns which suggest that during the first wave of the pandemic, there were only few variables that acted as strong and robust correlates with the final number of registered coronavirus infections and deaths in a country. These variables are related to the effect of density in social interactions and the overweight prevalence within the population. A simple correlation analysis indicates that the heterogeneity between the countries in terms of their health, social and economic nature might be the driver of this conclusion. Thus, the initial BMA results cannot capture (potentially) significant interactions between the correlates that are relevant to a particular country. To deal with this issue, we develop the coronavirus correlates Jointness Space. The Jointness Space models the interrelation between the potential correlates in explaining the coronavirus infections/deaths outcome, and can represent a statistical foundation for understanding the relationships between variables when developing policy recommendations for preventing future epidemic crises. Using this space, we find that the routes for reducing the potential negative impact of COVID-19 should focus on decreasing the prevalence of overweight people in the population and a small number of other variables that are relevant to that is studied. This will reduce both the registered infections and the observed deaths due to the COVID-19 disease. In the absence of realistic models that adequately cover all relevant aspects, this study provides the first step towards a more comprehensive understanding of the relationship between the socio-economic correlates of the coronavirus pandemic.

\section{Preliminaries}

\subsection{Measuring COVID-19 infections and death rates}
In a formal setting, the final number of registered COVID-19 infections per million population (p.m.p.) and the number of total COVID-19 deaths p.m.p. during the first wave of the pandemic are a result of a disease spreading process~\cite{wu2020nowcasting,kucharski2020early}. The extent to which a disease spreads within a population is uniquely determined by its reproduction number. This number describes the expected number of cases directly generated by one case in a population in which all individuals are susceptible to infection~\cite{bailey1975mathematical,van2008further}. Obviously, its magnitude depends on various natural characteristics of the disease, such as its infectivity or the duration of infectiousness, and the social distancing measures imposed by the government. Also, it depends on an abundance of health and socio-economic factors that govern the behavioral interactions within a population~\cite{keeling2011modeling,klepac2020contacts}. 

In general, we never observe the reproduction number, but rather the disease outcome, i.e., the number of infections/deaths. Thus, it is mathematically complex and computationally expensive to try and infer the reproduction number. To circumvent this problem, we utilize its known characteristics and derive a much simpler statistical model for the COVID-19 outcome. Here we choose a specific formulation where the disease outcome is modeled through the linear regression framework where the dependent variable is either the log of accumulated number of registered COVID-19 infections p.m.p. or the log of the accumulated number of COVID-19 deaths p.m.p. of the country at the end of the first wave of the pandemic. We focus on registered quantities normalized on per capita basis for the dependent variable instead of raw values to eliminate the bias in the outcomes arising from the different population sizes in the studied countries. The accumulation of the registered infections and deaths spans from the day of observation of the first infection in the country, up until the last day of the first wave of the pandemic in that country. The last day is, in general, different for each country and is inferred on the basis of the level of daily government response. The estimation procedure used to infer the last day of the first wave will be discussed in more detail in the next section.

The log transformation of the COVID-19 infections/deaths p.m.p. reduces the skewness of the original data and makes the dependent variable real-valued and continuous. For a such dependent variable, the linear regression framework is the simplest tool that quantifies the marginal effect of a set of potential independent variables (correlates). Its advantage lies in the efficient and unbiased analytical inference of the strength of the linear relationship. As such it has been widely used in modeling the outcome of epidemiological phenomena (See for example Refs.~\cite{wang2007obesity,fogli2012germs,carr2010systematic}).

A central question which arises in the model specification is the selection of the independent variables. While a
literature review can offer a comprehensive overview of all potential correlates, in reality we are never certain in their credibility. To reduce our uncertainty, we resort to BMA. BMA leverages Bayesian statistics to account for model uncertainty by estimating each possible
specification, and thus evaluating the posterior distribution of each parameter value and probability
that a particular model is the correct one~\cite{moral2015model}.This has allowed the BMA technique to be used in various domains, ranging from studying correlates of economic growth~\cite{moral2012determinants}, up to determinants of innovation processes~\cite{santa2019robust}. Recently, it was even applied for estimating the output losses during the Covid-19 pandemic~\cite{glocker2021determinants}.
\subsection{Baseline model}

The BMA method relies on the estimation of a baseline model that is used for evaluating the performance of all other models. In our case, this is the model which encompasses only variables for the state of the epidemic dynamics within the country and effect of government policies regarding social distancing, contact tracing and testing procedures.

We use two variables to quantify the possibility that countries are in a different state of the disease spreading process. The first variable simply measures the duration of epidemics in a country as the number of days since the first registered infection. In addition, we evaluate the time which the country had to prepare for the first wave of coronavirus. This is given as the number of days between the first registered infection worldwide and the first infection in the country. 

In order to assess the effect of government policies regarding social distancing and testing we construct an aggregated government response index. The index quantifies the average daily variation in government responses to the epidemic dynamics. As a measure for the daily variation, we take the Oxford COVID-19 government response index~\cite{hale2020variation}. The Oxford COVID-19 government response index is a composite measure that combines the daily effect of policies on social distancing, testing and contact tracing in an economy. For each country, we construct a weighted average of the index from all available data since their first registered coronavirus infection, up until the end date, i.e., the date when the government response index is at its maximum value. This threshold is chosen as a means to capture the moment when a country gains the ability to control and stabilize the propagation of the disease. To emphasize the effect of policy responses implemented on earlier dates, we construct a weighted average by putting a larger weight on those dates. This is because earlier responses are supposed to have a bigger impact on the prevention of the spread of the virus. The procedure implemented to derive the average government response index is described in Section S1 of the Supplementary Information (SI). 

Fig.~\ref{fig:1st-step} visualizes the results from the baseline model. We observe that the countries which had more detailed response policies also had less COVID-19 infections and mortality rates, as expected. In addition, the countries with longer duration of the crisis registered more infections and deaths p.m.p., whereas the countries which had more time to prepare for the crisis also had less infections and deaths.

It is apparent that the baseline model already has a large coefficient of determination ($R^2$) and can significantly explain a certain amount of the cross country variations in registered COVID infections/deaths p.m.p.. However, there is still a large amount of variation which, we conjecture that can be attributed to various health, social and economic correlates present within a society.


\begin{figure*}[t!]
\includegraphics[width=15cm]{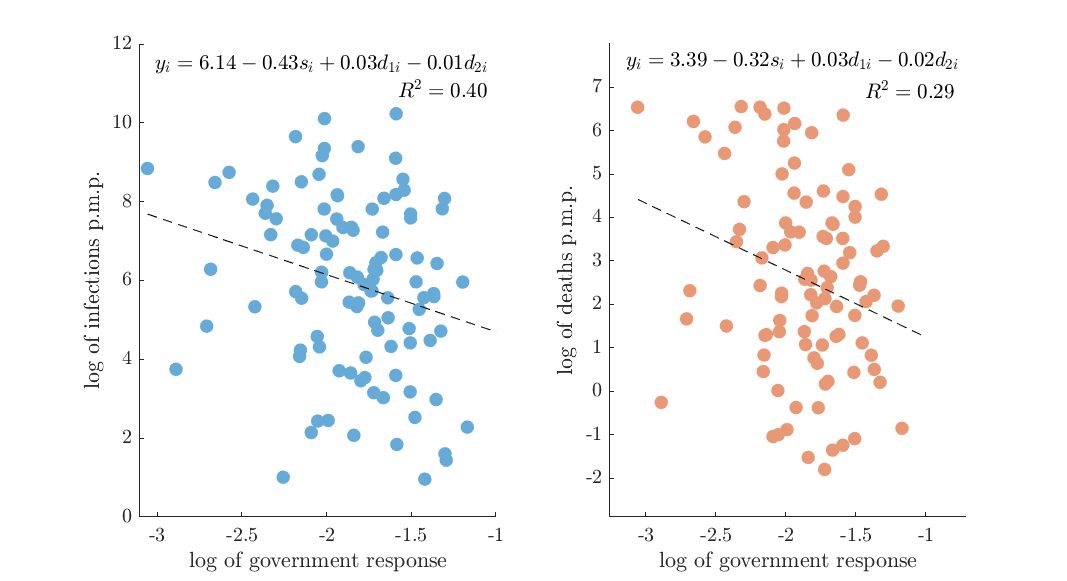}
\caption{\textbf{Explained variation in COVID-19 cases due to government response.}  \label{fig:1st-step}}
\end{figure*}

\subsection{Health, social and economic correlates}

To derive the set of potential health, social and economic correlates of the COVID-19 infection and mortality rates during the first wave of the pandemic we conduct a comprehensive literature review. From the literature review we recognize a total of 28 potential correlates, listed in Table~\ref{tab:list-determinants}. For a detailed description of the potential effect of the correlates we refer to the references given in the same table, and the references therein. We hereby point out that the data for each potential correlate corresponds to the last observed value (the value in 2019). This prevents the possible problem of endogenous independent variables in the specification of the regression.

In what follows, we only describe in short, the set of potential correlates on the basis of their characteristics. 

\paragraph{Healthcare Infrastructure:} The healthcare infrastructure essentially determines both the quantity and quality with which health care services are delivered in a time of an epidemic. As measures, we include 2 variables which capture the quantity of hospital beds, nurses and medical practitioners, as well as the quality of the coverage of essential health services.
On the one hand, studies report that well-structured healthcare resources positively affect a country's capacity to deal with epidemic emergencies~\cite{zanakis2007socio,itzwerth2006pandemic,whitley2006seasonal,breiman2007preparedness,adini2009relationship,garrett2009mitigating,oshitani2008major,gizelis2017maternal}. On the other hand, the healthcare infrastructure also greatly impacts the country's ability to perform testing and reporting when identifying the infected people. In this regard, economies with better structure are able to easily perform mass testing and more detailed reporting~\cite{hosseini2010predictive,quinn2014health,hogan2018monitoring}.

\paragraph{National health statistics:} The physical and mental state of a person play an important role in the degree to which the individual is susceptible to a disease. In countries where a significant proportion of the population suffer from diseases highly associated with the spread of an infectious disease as well as its fatal outcomes, we would expect more severe consequences of the emergent epidemics~\cite{marmot2005social,chen2008modelling,kelly2011scourge,nguyen2003epidemiology}. Specifically, metabolic disorders such as diabetes may intensify epidemic complications~\cite{susan2010global,allard2010diabetes}, whereas it has been observed that the susceptibility to various diseases account for the majority of deaths in complex emergencies~\cite{connolly2004communicable}. In addition, there is empirical evidence that adequate hygiene greatly reduces the rate of mortality, whereas overweight or asthma prevalence in the population may increase the fatality of epidemic diseases~\cite{abrams2020asthma,bassim2008modification,muller2015oral}. To quantify the national health characteristics, we include 6 variables that assess the general health level in the studied countries.

\paragraph{Economic performance:} We evaluate the economic performance of a country through 4 variables. This performance often mirrors the country's ability to intervene in a case of a public health crisis~\cite{strauss1998health,i2005health,sachs2001macroeconomics,ashraf2008does,wobst2004hiv,markowitz2010effects}. Variables such as GDP per capita have been used in modeling health outcomes, mortality trends, cause-specific mortality estimation and health system performance and finances~\cite{preston1975changing,james2012developing,nagano2020heart}. For poor countries, economic performance appears to improve health by providing the means to meet essential needs such as food, clean water and shelter, as well access to basic health care services. However, after a country reaches a certain threshold of development, few health benefits arise from further economic growth. It has been suggested that this is the reason why, contrary to expectations, the economic downturns during the 20th century were associated with declines in mortality rates~\cite{bezruchka2009effect,granados2008reversal}. Observations indicate that what drives the health in industrialized countries is not absolute wealth or growth but how the nation’s resources are shared across the population~\cite{wilkinson2010spirit}. The more egalitarian income distribution within a rich country is associated with better health of population~\cite{ezzati2008reversal,siddiqi2007towards,kawachi1999income,krisberg2016income}. 

\paragraph{Societal characteristics:} The characteristics of a society often reveal the way in which people interact, and thus spread the disease. In this aspect, properties such as education and the degree of digitalization within a society reflect the level of a person's reaction and promotion of self-induced measures for reducing the spread of the disease~\cite{putnam2001social,folland2007does,lee2013comparative,baker2011education,mackenbach2008socioeconomic}. Also, the way we mix in society may effectively control the spread of infectious diseases~\cite{hens2009estimating,mossong2008social,melegaro2011types,prem2017projecting,klepac2020contacts}. 
To measure the societal characteristics, we identify 4 variables. 

\paragraph{Demographic structure:} Similarly, to the national health statistics, the demographic structure may impact the average susceptibility of the population to a disease. Certain demographic groups may simply have weaker defensive health mechanisms to cope with the stress induced by the disease~\cite{wallinga2006using,erkoreka2010spanish,armstrong1999trends,ainsworth2003impact}. In addition, the location of living may greatly affect the way in which the disease is spread~\cite{mastrandrea2015contact,kucharski2014contribution}. To express these phenomena, we collect 7 variables.

\paragraph{Natural environment:}  Numerous studies discuss possible correlation between air pollution and COVID-19 infections and mortality rates~\cite{braga2000respiratory,wu2020exposure,clay2018pollution}. In addition, some authors note that countries where natural sustainability is deteriorated, are also more vulnerable to epidemic outbreak~\cite{di2020opinion}. On the other hand, healthy natural environments may attract more tourists, which could drive the disease spread~\cite{hosseini2010predictive}. Finally, weather patterns can also impact the infectiousness of the disease, especially exposure when there are very cold days in winter and very hot days in summer~\cite{simiao2021climate2}. We gather the data for 5 variables which capture the essence of this characteristic.

\begin{table}[H]
\caption{\textbf{List of Potential correlates of the COVID-19 first wave infections and mortality rates.}\label{tab:list-determinants}}
\resizebox{\textwidth}{!}{%
\begin{tabular}{|l|l|c|l|}
\hline
\multicolumn{1}{|c|}{\textbf{Variable}} & \multicolumn{1}{|c|}{\textbf{Measure}} & \multicolumn{1}{|c|}{\textbf{Source}} & \multicolumn{1}{|c|}{\textbf{Refs.}} \\\hline\hline
\rowcolor{table-color-dark}\multicolumn{4}{|c|}{\textbf{Healthcare Infrastructure}} \\\hline
\rowcolor{table-color-light} Medical resources & Medical resources index & WDI & ~\cite{hosseini2010predictive,zanakis2007socio,whitley2006seasonal,breiman2007preparedness,adini2009relationship,garrett2009mitigating,itzwerth2006pandemic,oshitani2008major,hogan2018monitoring,quinn2014health,gizelis2017maternal} \\
\rowcolor{table-color-light}Health coverage&UHC service coverage index&WDI& ~\cite{hosseini2010predictive,zanakis2007socio,whitley2006seasonal,breiman2007preparedness,adini2009relationship,garrett2009mitigating,itzwerth2006pandemic,oshitani2008major,hogan2018monitoring,quinn2014health,gizelis2017maternal} \\\hline
\rowcolor{table-color-dark} \rowcolor{table-color-dark}\multicolumn{4}{|c|}{\textbf{National health statistics}} \\\hline
\rowcolor{table-color-light}Life expectancy&Life expectancy at birth &WDI&\cite{marmot2005social,chen2008modelling,kelly2011scourge,nguyen2003epidemiology}  \\
\rowcolor{table-color-light}Mortality & Non-natural causes mortality  index &WDI&~\cite{susan2010global,allard2010diabetes,connolly2004communicable,bassim2008modification,muller2015oral} \\
\rowcolor{table-color-light} Comorbidities & Comorbidities index &Our World in Data&~\cite{susan2010global,allard2010diabetes,connolly2004communicable,bassim2008modification,muller2015oral} \\
\rowcolor{table-color-light}Immunization & Immunization index &WDI&~\cite{zanakis2007socio} \\
\rowcolor{table-color-light} Overweight prevalence & \% of adults with BMI $>$ 25 kg/m2 & ESG &\cite{lighter2020obesity,sattar2020obesity,stefan2020obesity} \\
\rowcolor{table-color-light} Asthma prevalence & \% of population with Asthma & Our World in Data &\cite{abrams2020asthma} \\\hline
\rowcolor{table-color-dark}\multicolumn{4}{|c|}{\textbf{Economic performance}} \\\hline
\rowcolor{table-color-light}Economic development&GDP p.c., PPP \$&WDI& ~\cite{strauss1998health,i2005health,sachs2001macroeconomics,ashraf2008does,preston1975changing,james2012developing,nagano2020heart} \\
\rowcolor{table-color-light}Labor market&Employment to population ratio&WDI& \cite{strauss1998health,zanakis2007socio,wobst2004hiv,markowitz2010effects} \\
\rowcolor{table-color-light}Government spending&Gov. health spending p.c., PPP \$&WDI&\cite{hosseini2010predictive,strauss1998health,i2005health,sachs2001macroeconomics,ashraf2008does} \\
\rowcolor{table-color-light}Income inequality&GINI index&WDI&\cite{wilkinson2010spirit,ezzati2008reversal,siddiqi2007towards,kawachi1999income,krisberg2016income} \\
\hline
\rowcolor{table-color-dark}\multicolumn{4}{|c|}{\textbf{Societal characteristics}} \\\hline
\rowcolor{table-color-light}Social connectedness &Social connectedness index (PageRank) & DFG&\cite{bailey2018social,kuchler2020geographic} \\
\rowcolor{table-color-light}Digitalization &  Digitalization index & WDI&\cite{zanakis2007socio,putnam2001social,folland2007does,lee2013comparative,baker2011education,mackenbach2008socioeconomic} \\
\rowcolor{table-color-light}Education& Human capital index& WDI&\cite{marmot2005social,putnam2001social,folland2007does,lee2013comparative,baker2011education,mackenbach2008socioeconomic}  \\
\rowcolor{table-color-light}Household size& Avg. no. of persons in a household& UN& ~\cite{hens2009estimating,mossong2008social,melegaro2011types,prem2017projecting,klepac2020contacts} \\
\hline
\rowcolor{table-color-dark}\multicolumn{4}{|c|}{\textbf{Demographic structure}} \\\hline
\rowcolor{table-color-light}Elderly population&Population age 65+ (\% of total)&WDI&\cite{wallinga2006using,erkoreka2010spanish,armstrong1999trends,ainsworth2003impact} \\
\rowcolor{table-color-light}Young population&Population ages 0-14 (\% of total)&WDI&\cite{wallinga2006using,erkoreka2010spanish,armstrong1999trends,ainsworth2003impact} \\
\rowcolor{table-color-light}Gender &50\%+ male population (\% of total) &WDI&\cite{wallinga2006using,erkoreka2010spanish,armstrong1999trends,ainsworth2003impact} \\
\rowcolor{table-color-light}Population size&Population, total&WM & \cite{mastrandrea2015contact,kucharski2014contribution} \\
\rowcolor{table-color-light}Rural population&Rural population (\% of total)&WDI&\cite{mastrandrea2015contact,kucharski2014contribution} \\
\rowcolor{table-color-light}Migration&Int. migrant stock (\% of population)&WDI& \cite{mastrandrea2015contact,kucharski2014contribution} \\
\rowcolor{table-color-light}Population density& People per sq. km &WDI& \cite{mastrandrea2015contact,kucharski2014contribution} \\\hline
\rowcolor{table-color-dark}\multicolumn{4}{|c|}{\textbf{Natural environment}} \\\hline
\rowcolor{table-color-light}Sustainable development& Ecological Footprint (gha/person)&GFN&\cite{di2020opinion}  \\
\rowcolor{table-color-light}Air Pollution&Yearly avg P.M. 2.5 exposure&SGA&\cite{braga2000respiratory,wu2020exposure,clay2018pollution} \\
\rowcolor{table-color-light}Weather (latitude)& Geographic coordinate: Latitude&Google&\cite{simiao2021climate2} \\
\rowcolor{table-color-light}Air transport&Yearly passengers carried&WDI&\cite{hosseini2010predictive} \\
\rowcolor{table-color-light}International Tourism& Number of tourist arrivals&WDI&\cite{hosseini2010predictive} \\
\hline
\end{tabular}}
\end{table}

\section{Results}

\subsection{BMA estimation} 

We use this set of variables and estimate two distinct BMA models. In the first model the dependent variable is the log of COVID-19 infections p.m.p., whereas in the second model we investigate the critical correlates of the log of the mortality rate due to the coronavirus. For the estimation procedure we use data on 105 countries. This is the maximal set of countries for which the data on all 28 potential correlates could be attained. The summary statistics and the data gathering and preprocessing procedures are described in SI Section \ref{sec:data}. The mathematical background of BMA together with our inference setup is given in SI Section \ref{sec:BMA}.

\begin{figure}[t!]
\begin{subfigure}{.5\textwidth}
  \centering
  \includegraphics[width=\linewidth]{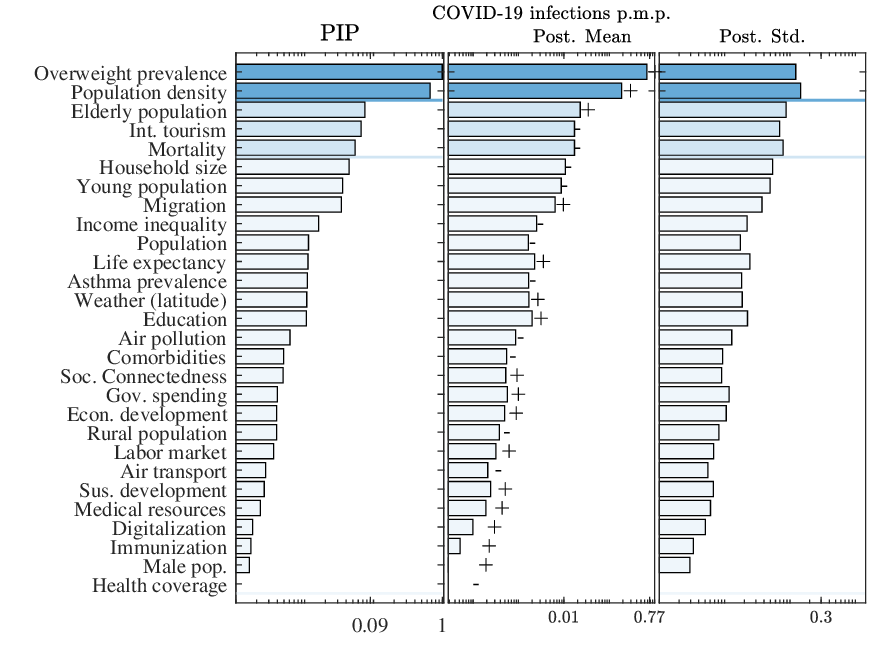}
\end{subfigure}%
\begin{subfigure}{.5\textwidth}
  \centering
  \includegraphics[width=\linewidth]{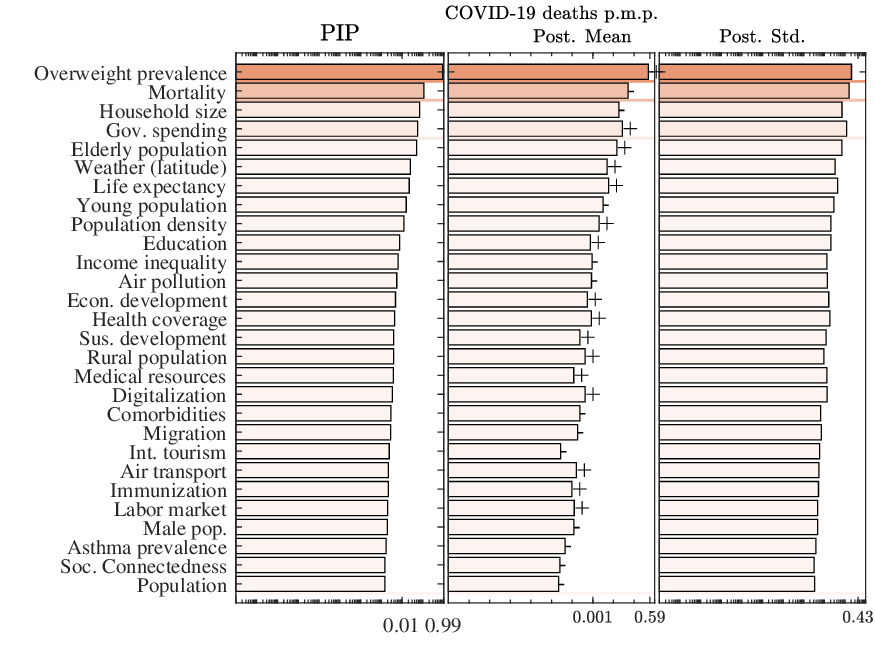}
\end{subfigure}
\caption{\textbf{BMA results.} Bars for the posterior inclusion probability (PIP), posterior mean (Post. Mean) and the posterior standard deviation (Post. Std.) of each potential correlate. The variables are ordered according to their PIP. The Post. Mean is in absolute value. The signs next to the bar of each variable indicate the direction of its impact. The horizontal lines divide the variables into groups according to their PIP. The horizontal axis is on a logarithmic scale. The setup used to estimate the results is described in SI Section \ref{sec:BMA}.}
\label{fig:BMA-results}
\end{figure}

Fig.~\ref{fig:BMA-results} displays the respective results. In both situations, the variables are ordered according to their posterior inclusion probabilities (PIP), given in the second column. PIP quantifies the posterior probability that a given correlate belongs to the linear regression model that best describes the COVID-19 infections/mortality rates. Besides this statistic, we also provide the posterior mean (Post mean) and the posterior standard deviation (Post Std). Post mean is an estimate of the average magnitude of the effect of a correlate, whereas the Post Std evaluates the deviation from this value.

In the inference procedure (described in SI Section \ref{sec:BMA}) we initially assumed that the linear regression model which best describes the COVID-19 first wave infections and mortality rates is a result of the baseline specification and 3 additional variables. Our prior belief stems from the general observation which suggests that economies are heterogeneous, and a small number of complementing factors may contribute to the extent of the coronavirus spread, while the other potential correlates may simply behave as substitutes in terms of socio-economic interpretation within a country. Nevertheless, we found that our results do not depend on the prior assumption of the size of the true model. Altogether, this implies that the prior inclusion probability of each potential correlate is around $0.1$. We use this attribute, together with the posterior inclusion probability of each correlate, to divide the correlates into four disjoint groups:

\paragraph{Correlates with strong evidence:} (PIP $> 0.5$). The first group describes the correlates which have by far larger posterior inclusion probability than prior probability, and thus there is strong evidence to be included in the true model. We find two such variables related to explaining the coronavirus infections, the overweight prevalence in the country and the population density. Both variables are positively related with the number of registered COVID-19 infections p.m.p.. When investigating the critical correlates of the coronavirus deaths, it appears that the overweight prevalence is the only variable for which there is strong evidence to explain the outcome and has a positive impact.

\paragraph{Correlates with medium evidence:} ($0.5 \geq$ PIP $> 0.1$). There are no variables for which there is medium evidence to be a correlate of the COVID-19 number of infections in the first wave, whereas mortality from non-natural causes is the only variable for which there is medium evidence to be a correlate of the COVID-19 death rate, with a negative effect. 

\paragraph{Correlates with weak evidence:} ($0.1\geq$PIP$>0.05$). These are correlates which have lower posterior inclusion probability than their prior one, but still may account for some of the variations in the COVID-19 infections/deaths. For the infections per million population there are three such correlates, the fraction of elderly population, the number of international tourist arrivals and the mortality from non-natural causes. The elderly population has a positive Post Mean, whereas the other two variables have negative Post Mean. When studying the COVID-19 death rate, we find two correlates with weak evidence. They are the household size and the government health expenditure. The household size has a positive marginal effect (Post Mean), whereas the government health expenditure shows a negative effect.
 
\paragraph{Correlates with negligible evidence:} (PIP$\leq0.05$). All other variables have negligible evidence to be a true correlate of the coronavirus outcome. In total, we find negligible evidence for explaining the coronavirus infections in 23 variables and for explaining the coronavirus deaths in 24 variables.

The division of the variables into groups allows us to assess the robustness of each potential correlate -- those belonging to a group described with a larger PIP also offer more credible explanation for the coronavirus infections and death rates. Nonetheless, we point out that although the comparison between posterior inclusion probabilities and prior inclusion probabilities is a common approach, its interpretation must be taken with care. Concretely, the inhomogeneous nature of the specific features of the countries can drive our results. The presence of this phenomenon in our data be inferred by conducting a simple correlation analysis between the potential correlates. If the variables are highly correlated between each other then there is a problem of multicolinearity. Multicolinearity can lead to wider credible intervals that eventually produce less statistically reliable posterior inclusion probabilities in terms of the effect of independent variables in a model. As said in~\cite{moral2012determinants}, even if the posterior inclusion probability is lower than the prior inclusion probability for a given variable, it might be that this particular variable is important to decision makers under certain circumstances.

In SI Section \ref{sec:BMA-robustness} we conduct several checks to confirm the robustness of our results. In the first robustness check we investigate the impact of outliers. In particular, definitely there were several countries which were either extremely affected by the coronavirus or displayed great immunity to the epidemic crisis. To check the robustness of our results against the presence of such data we implement the following strategy. First, we remove a country from the sample. Then, we re-perform the BMA procedure with the resulting countries. We repeat this procedure for every country and recover the median results for each potential correlate. The results indicate that the findings presented here are valid even in the presence of outliers. In the same section, we display the economies which contributed most and least to the credibility of a particular variable. These are the countries which, when excluded, lead to the minimum, respectively maximum, posterior inclusion probability of the given variable. The investigation suggests that there are multiple countries which are significant contributors to the PIP value of each correlate, thus further indicating that there is heterogeneity in the health social and economic features of the countries. In the second check, we change the end date of the pandemic to be equal to the first date after the day at which the daily government response index is at its maximum and that is at least 20\% lower than the daily maximum. This effectively prolongs the duration of the first wave. Nonetheless, it still does not impact the findings. In the third check, we change the dependent variable to be the raw number of infections and deaths at the end of the first wave. In other words, now the dependent variable describes counts and the linear regression framework is not a suitable model. Instead, for the estimation of the marginal impact we use a quasi-Poisson model, which is the most often used procedure when the dependent variable is given as a count that has a large variance~\cite{ver2007quasi}. Even in this case, the results do not change. In the final robustness check, we add a spatial weighting matrix in the baseline model in order to account for the potential spatial autocorrelation in the spread of COVID-19. Multiple studies have indicated that this effect might exist (See for example~\cite{krisztin2020spatial}). Again our findings do not significantly change.

Definitely, even if useful for presentation purposes, the mechanical application of a threshold, or a simple comparison between the prior and the posterior, should often be avoided in practice. Each BMA analysis should be coupled with an investigation for the interrelationships between the variables in explaining the dependent variable. We perform this analysis in the subsequent section.

\subsection{``Jointness Space'' of the COVID-19 infections/deaths correlates}

The next step in deriving the linear regression model that describes best the coronavirus infections/mortality rates is to find its dimension, i.e., the number of explanatory variables included in the model. As a measure for this quantity, BMA provides the posterior size, formally defined as the posterior belief for the dimension of the model. We find that, for the coronavirus infections p.m.p. the posterior model size is $2.21$ whereas for the coronavirus deaths p.m.p. it is $1.34$.

After discovering the model size, we need to specify the explanatory variables. This raises the issue of how to construct the appropriate model. One possible solution is to use the correlates with the highest PIP value and regress them on the dependent variable. However, this neglects the interdependence of inclusion and exclusion of correlates in a same model. A standard approach for resolving this issue is to conduct a statistical \textit{jointness} test. The concept of jointness has been introduced within the BMA framework with the aim to capture dependence between explanatory variables in the posterior distribution over the model space~\cite{doppelhofer2009jointness}. By emphasizing dependence and conditioning on a set of one or more other variables, jointness moves away from marginal measures of variable importance and investigates the sensitivity of posterior distributions of parameters of interest to dependence across regressors. For example, if two variables are complementary in their posterior distribution over the model space, models that either include or exclude both variables together receive relatively more weight than models where only one variable is present. In our context, jointness tests will allow us to infer whether two variables are complements, i.e., tend to be included together in models with high posterior probability, or substitutes, i.e., models with high posterior probability tend to exclude the joint inclusion of both variables. 

To better understand the properties of the COVID-19 infection and mortality rates during the first wave, we perform the jointness test developed by Hofmarcher et al.~\cite{hofmarcher2018bivariate}. Using this test we can estimate a metric between each pair of correlates and quantify their relationship in a range between $-1$ and $1$. In the two extremes, $-1$ indicates that the two correlates behave as perfect substitutes in the true model, whereas $1$ indicates that they are included in the true model together. The resulting jointness metric between pairs of correlates can be used to construct a network (graph),  which we refer to as the \textit{Jointness Space} of the COVID-19 correlates. In this network, the nodes are the potential health, social and economic correlates, whereas the jointness values represent the edge weights. In other words, two arbitrary correlates are linked with each other by the posterior belief that both of them belong to the same linear regression model governing the coronavirus infections/mortality rate. 

In theory, many possible factors may cause complementarity between the variables, such as national culture~\cite{meeuwesen2009can}, the type of healthcare system~\cite{mossialos20162015} or political priorities~\cite{boas2019electoral}. All of these are a priori notions of what dimension drives the relatedness between the potential correlates and assume that there is little flexibility in choosing the correct model. Instead, the jointness space follows an agnostic approach and uses a data-driven measure, based on the idea that, if two correlates are related because they offer contrasting information regarding the coronavirus outcome they will tend to be included in the true model in tandem, whereas variables that give similar information are less likely to be included together. Hence, the developed network offers a statistical view for the importance of the social, health and economic correlates when developing policies aimed at reducing the impact of epidemic crises.

\begin{figure*}[t!]
\includegraphics[width=15cm]{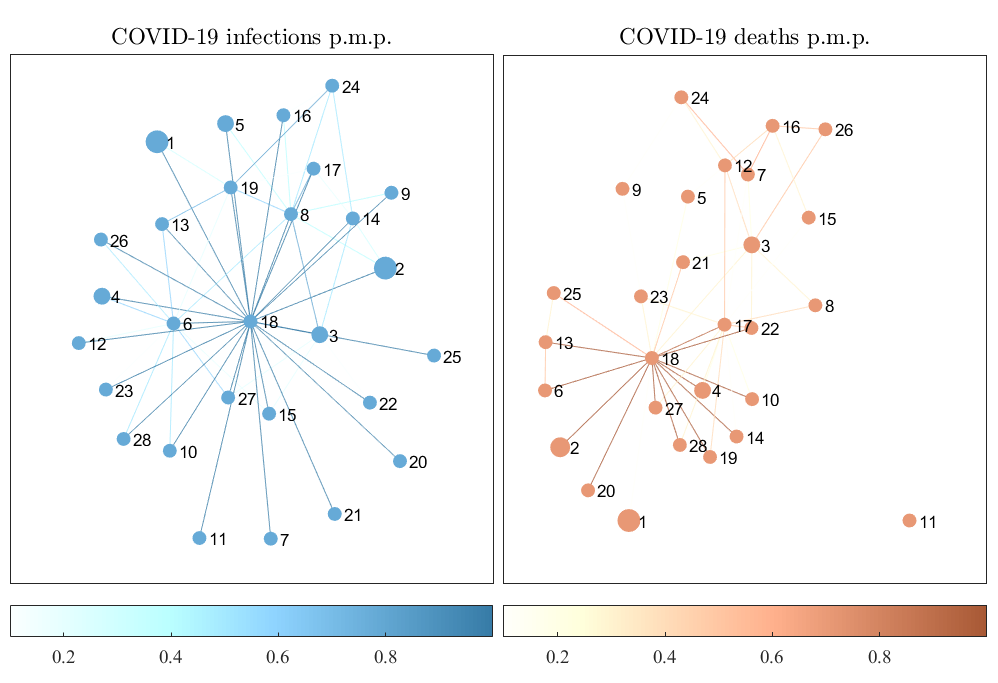}
\resizebox{\textwidth}{!}{
\begin{tabular}{|d{0}|l|d{0}|l|d{0}|l|d{0}|l|d{0}|l|d{0}|l|}\hline
\multicolumn{12}{|c|}{\textbf{Node labels}}\\\hline
 \rowcolor{table-color-light}1.& Life expectancy           &  6.& Population density & 11.& Health coverage &  16.& Gov.h.spending &  21.& Digitalization & 26.& Comorbidities \\
 \rowcolor{table-color-light}2.&  Young population &  7.& Rural population &  12.& Migration &  17.& Household size &  22.&Medical resources& 27.  & Soc. Connectedness\\
 \rowcolor{table-color-light}3.&  Elderly population &  8.& Int. tourism  &  13.& Air pollution  &  18.& Overweight prevalence &  23.& Immunization&   28.& Gender\\
 \rowcolor{table-color-light}4.&  Population        & 9.& Education & 14.& Air transport &  19.& Asthma prevalence &  24.& Mortality& & \\
 \rowcolor{table-color-light}5.&  Econ. development   & 10.& Labor market  & 15.& Sus. development &  20.& Income inequality &  25.& Weather&  & \\   
 \rowcolor{table-color-light}&  & & & & &  & &  &  &  &
   \\\hline 
\end{tabular}}
\caption{\textbf{Jointness Space of the COVID-19 correlates.} The color of the edge between a pair of correlates is proportional to their Jointness metric. To visualize the network, we use the Force-Layout drawing algorithm. \label{fig:jointness}}
\end{figure*}

The networks depicted in Fig.~\ref{fig:jointness} visualize the Jointness Space of the correlates included in our BMA framework. To emphasize the complementary relationships, we connect only correlates with positive jointness. The full description for the procedure implemented for constructing the Jointness Space is given in SI Section \ref{sec:jointness}. In the networks, the correlates which can be included in multiple models take a more central position whereas the periphery is constituted of correlates whose credibility in explaining the coronavirus outcome mostly substitutes the effect of other variables. 

Interestingly, we observe that the topological form of the Jointness Space is not significantly determined by how we specify the dependent variable. In both situations, there is one large connected component with correlates where the central role is played by the overweight prevalence. Thus, the obtained maps suggest the first step in the construction of the linear-regression model for the COVID-19 infections/death rate in the first wave is by first focusing on the fraction of overweight persons in the country. Moreover, almost all other variables belong to the same component. Only in the case when the dependent variable is modelled through the COVID-19 deaths, Life expectancy and Health coverage are excluded from the component. Hence, the variables included in our analysis are complements in explaining the COVID-19 infections/death rates. Based on this finding, we once again assert that the next variables that will be included in the model, should be specific for the economy that is the subject of the study. Nonetheless, improving the features of the correlates that are located more centrally might yield a synergistic effect, thus significantly reducing the risk of a more negative COVID-19 infections/death rate.

\section{Conclusion and discussion}

In this work, we utilized Bayesian model averaging techniques to provide a comprehensive analysis for the health, social and economic correlates of that contributed to between country differences in the final number of infections and deaths during the first wave of the COVID-19 pandemic. Our findings suggest that government response policies, such as testing procedures, tracking of individuals and social distancing measures, and the state of the dynamics of the disease spread can significantly explain the variety in the coronavirus outcome between the countries. Aside from these variables, only a handful of additional variables are able to robustly explain the extent of the COVID-19 infection/deaths and thus provide general rules for the virus spread.

The sole variable strongly related to  the coronavirus deaths is the overweight prevalence. Countries with a larger fraction of overweight population also show greater susceptibility to fatal virus outcomes. Interestingly, besides the overweight prevalence, the population density is also a strong correlate of the registered coronavirus infections per million population. More densely populated countries display higher infection rates. A plentiful of reasons can be used as a possible interpretation for these results. For instance, it is known that the degree of disease spread scales proportionally with population density~\cite{draief2006thresholds}. This is because, everything else considered, in denser populations typically there is more social mixing~\cite{klepac2020contacts}. In a similar fashion, various explanations can be found for the observed effect of overweight prevalence. In particular, the prevalence of overweight people is closely related to unhealthy habits of living and, hence, larger susceptibility to both disease infections and fatal outcomes.

The robustness checks and the performed jointness analysis suggested that the insignificance of the other variables might not be the reason for their low PIP values. Instead, the variables which we studied have complementary effect in explaining the COVID-19 infections and death rates of the first wave of the pandemic. This lead us to suspect that the results are driven by the heterogeneous health, social and economic features of the countries. To this end, an interesting topic for future research would be to explore how the effect of the correlates evolved during the different waves of the pandemic. In the absence of a unifying framework covering the relevant aspects of the interrelation between the potential correlates during the various waves, the jointness analysis performed here (and the resulting Jointness Space) can provide the starting point for the development of a more comprehensive understanding of the factors determining the infection and mortality rates of the pandemic. Moreover, with an improved understanding of the dynamics of the coronavirus pandemic, the insights obtained from this analysis can influence the development of appropriate policy recommendations.




\newpage
\section*{Supplementary information}

\setcounter{equation}{0}
\setcounter{figure}{0}
\setcounter{table}{0}
\setcounter{theorem}{0}
\setcounter{subsection}{0}
\setcounter{subsubsection}{0}
\makeatletter
\renewcommand{\theequation}{S\arabic{equation}}
\renewcommand{\thesubsection}{S\arabic{subsection}}
\renewcommand{\thetable}{S\arabic{table}}
\renewcommand{\thefigure}{S\arabic{figure}}
\renewcommand{\thetheorem}{S\arabic{theorem}}
\renewcommand{\theproposition}{S\arabic{proposition}}

\subsection{Calculation of the Government response index}
\label{sec:stringency-index}

To calculate our government response measure, we make use of Oxford's daily government response index. Oxford's daily government response index measures, on a scale of 1-100, the variation in daily government responses to COVID-19 by accumulating ordinal data on country social distancing measures on school, workplace and public transport closure; cancellation of public events; restrictions of internal movement; control of international travel and promotion of public campaigns on prevention of coronavirus spread; testing policies and procedures implemented for tracing contacts of infected individuals. We refer to~\cite{hale2020variation} for a detailed overview on how the daily index is constructed.

To calculate the overall government response index $c_i(d_i^*)$ at the final date $d_i^*$ from the provided daily indexes we implement the following procedure. Let $C_i(t)$ represent the government response on day $t$, where $t = 1,2 \dots d_i^*$, then our index can be estimated as
\begin{align}
    c_i(d_i^*) &= \sum_{s=1}^{d_i^*}w_i(s) C_i(s),
\end{align}
where $w_i(s)$ are the weights given to each day since the first registered case. We use a simple inverse weight procedure by giving larger weights to earlier dates, i.e.,
\begin{align}
    w_i(s) &= \frac{1}{s} / \sum_{k=1}^{d_i^*} \frac{1}{k}.
\end{align}

We choose the last date $d_i^*$ to be the last day at which the daily government response index $C_i(t)$ is at its maximum value.

\subsection{Data description}
\label{sec:data}

The data for the dependent variables are taken from Our World in Data coronavirus tracker. The tracker offers daily coverage of country coronavirus statistics, by collecting data mainly from the European Centre for Disease Prevention and Control. Because national aggregates often lag behind the regional and local health departments' data, an important part of the data collection process consists in utilizing thousands of daily reports released by local authorities. The results were made with data gathered on 13th November 2020.

The data used for measuring the possible health, social and economic correlates are gathered from 9 various sources. In particular, the collection is as follows: 19 variables are from the World Bank's World Development Indicators (WDI), 2 variables are from the Our World in Data database and there is 1 variable from World Bank's Environmental, social and governance data (ESG), the Worldometers database (WM), Data For Good (DFG), the State of Global Air (SGA), the Global footprint network (GFN),  United Nations (UN) database and Google. Six of the potential correlates were constructed by deriving our own index with data taken from the described source. The construction procedures for each of these variables are described in the following subsection. The full list of sources together with links to their websites is given in Table~\ref{tab:data-sources}. The data used in the analysis are available at \url{https://github.com/pero-jolak/coronavirus-socio-economic-determinants}.

\begin{table}[H]
\caption{\textbf{List of data sources.} \label{tab:data-sources}}
\resizebox{0.95\textwidth}{!}{%
\begin{tabular}{|c|l|}
\hline
\textbf{Source} & \textbf{Link}     \\\hline\hline
\rowcolor{table-color-light}COVID-19 infections/deaths & \url{ourworldindata.org/coronavirus} \\\hline
\rowcolor{table-color-light} DFG & \url{dataforgood.fb.com} \\\hline
\rowcolor{table-color-light} Google & \url{maps.google.com} \\\hline
\rowcolor{table-color-light} ESG & \url{datacatalog.worldbank.org/dataset/environment-social-and-governance-data} \\\hline
\rowcolor{table-color-light}GFN & \url{data.footprintnetwork.org}         \\\hline
\rowcolor{table-color-light}Gov. Response    & \url{covidtracker.bsg.ox.ac.uk}   \\\hline
\rowcolor{table-color-light}Our world in data                 & \url{ourworldindata.org}   \\\hline
\rowcolor{table-color-light}SGA                & \url{www.stateofglobalair.org/engage}   \\\hline
\rowcolor{table-color-light}UN                 & \url{data.un.org}               \\\hline
\rowcolor{table-color-light}WDI                & \url{data.worldbank.org/}   \\\hline
\rowcolor{table-color-light}WGI                & \url{info.worldbank.org/governance/wgi} \\\hline
\rowcolor{table-color-light}WM                & \url{/www.worldometers.info/world-population} \\\hline
\end{tabular}}
\end{table}

To reduce the noise from the data we, use only data for countries with population above 1 million. In addition, we only use countries for which there is data on all of the potential correlates. Table~\ref{tab:countries} gives the countries for which all of these data was available.

\begin{table}[H]
\caption{\textbf{List of countries and estimation period.} \label{tab:countries}}
\resizebox{0.95\textwidth}{!}{%
\begin{tabular}{|l|c|c||l|c|c||l|c|c|}
\hline
\rowcolor{table-color-dark}\textbf{Country}            & \textbf{First Date} & \textbf{End Date} & \textbf{Country}        & \textbf{First Date} & \textbf{End Date} & \textbf{Country}      & \textbf{First Date} & \textbf{End Date} \\ \hline
\rowcolor{table-color-light}Albania                & 09-Mar     & 31-May   & Georgia    & 27-Feb     & 26-Apr   & Nigeria             & 28-Feb     & 03-May   \\
\rowcolor{table-color-light}Argentina              & 04-Mar     & 25-Apr   & Ghana      & 13-Mar     & 17-Apr   & Netherlands         & 28-Feb     & 10-May   \\
\rowcolor{table-color-light}Australia              & 25-Jan     & 27-Aug   & Greece     & 27-Feb     & 13-Sep   & Norway              & 27-Feb     & 19-Apr   \\
\rowcolor{table-color-light}Austria                & 26-Feb     & 13-Apr   & Guatemala  & 15-Mar     & 26-Jul   & Nepal               & 25-Jan     & 18-Aug   \\
\rowcolor{table-color-light}Azerbaijan             & 29-Feb     & 05-Aug   & Honduras   & 12-Mar     & 07-Jun   & New Zealand         & 28-Feb     & 27-Apr   \\
\rowcolor{table-color-light}Belgium                & 04-Feb     & 04-May   & Croatia    & 26-Feb     & 26-Apr   & Pakistan            & 27-Feb     & 03-Jun   \\
\rowcolor{table-color-light}Benin                  & 17-Mar     & 10-May   & Hungary    & 05-Mar     & 03-May   & Panama              & 10-Mar     & 11-Oct   \\
\rowcolor{table-color-light}Burkina Faso           & 11-Mar     & 04-May   & Indonesia  & 02-Mar     & 02-May   & Peru                & 07-Mar     & 10-May   \\
\rowcolor{table-color-light}Bangladesh             & 09-Mar     & 30-May   & India      & 30-Jan     & 19-Apr   & Philippines         & 30-Jan     & 30-Apr   \\
\rowcolor{table-color-light}Bulgaria               & 08-Mar     & 30-Apr   & Ireland    & 01-Mar     & 29-Oct   & Papua New Guinea    & 21-Mar     & 11-Aug   \\
\rowcolor{table-color-light}Bosnia and Herzegovina & 06-Mar     & 23-Apr   & Iraq       & 25-Feb     & 26-Aug   & Poland              & 04-Mar     & 01-Nov   \\
\rowcolor{table-color-light}Bolivia                & 12-Mar     & 23-Jun   & Israel     & 22-Feb     & 16-Apr   & Portugal            & 03-Mar     & 03-May   \\
\rowcolor{table-color-light}Brazil                 & 26-Feb     & 28-Jul   & Italy      & 31-Jan     & 03-May   & Paraguay            & 08-Mar     & 24-May   \\
\rowcolor{table-color-light}Botswana               & 01-Apr     & 07-May   & Jamaica    & 12-Mar     & 30-May   & Romania             & 27-Feb     & 10-May   \\
\rowcolor{table-color-light}Canada                 & 26-Jan     & 11-Aug   & Jordan     & 03-Mar     & 20-Apr   & Russia              & 01-Feb     & 31-May   \\
\rowcolor{table-color-light}Switzerland            & 26-Feb     & 29-May   & Japan      & 15-Jan     & 13-May   & Rwanda              & 15-Mar     & 03-May   \\
\rowcolor{table-color-light}Chile                  & 04-Mar     & 04-Oct   & Kazakhstan & 15-Mar     & 10-May   & Senegal             & 03-Mar     & 10-May   \\
\rowcolor{table-color-light}Côte d'Ivoire          & 12-Mar     & 07-May   & Kenya      & 14-Mar     & 22-Jun   & Singapore           & 24-Jan     & 01-Jun   \\
\rowcolor{table-color-light}Cameroon               & 07-Mar     & 30-Apr   & Kyrgyzstan & 19-Mar     & 29-Apr   & El Salvador         & 19-Mar     & 01-Jun   \\
\rowcolor{table-color-light}Colombia               & 07-Mar     & 05-May   & Korea      & 20-Jan     & 17-Apr   & Serbia              & 07-Mar     & 20-Apr   \\
\rowcolor{table-color-light}Costa Rica             & 07-Mar     & 30-Apr   & LAOS       & 25-Mar     & 03-May   & Slovakia            & 07-Mar     & 29-Oct   \\
\rowcolor{table-color-light}Cyprus                 & 10-Mar     & 03-May   & Lithuania  & 28-Feb     & 13-Apr   & Slovenia            & 05-Mar     & 19-Apr   \\
\rowcolor{table-color-light}Czechia                & 02-Mar     & 01-Apr   & Latvia     & 03-Mar     & 11-May   & Sweden              & 01-Feb     & 12-Jun   \\
\rowcolor{table-color-light}Germany                & 28-Jan     & 02-May   & Morocco    & 03-Mar     & 10-Jun   & Togo                & 07-Mar     & 07-Jun   \\
\rowcolor{table-color-light}Denmark                & 27-Feb     & 21-May   & Moldova    & 08-Mar     & 15-May   & Thailand            & 13-Jan     & 02-May   \\
\rowcolor{table-color-light}Dominican Republic     & 02-Mar     & 17-May   & Madagascar & 21-Mar     & 19-Apr   & Trinidad and Tobago & 13-Mar     & 30-Apr   \\
\rowcolor{table-color-light}Ecuador                & 01-Mar     & 03-May   & Mexico     & 14-Jan     & 26-Oct   & Turkey              & 12-Mar     & 20-Sep   \\
\rowcolor{table-color-light}Egypt                  & 15-Feb     & 06-Jun   & Myanmar    & 18-Mar     & 24-Oct   & Tanzania            & 17-Mar     & 17-May   \\
\rowcolor{table-color-light}Spain                  & 01-Feb     & 16-May   & Mongolia   & 10-Mar     & 07-May   & Uganda              & 22-Mar     & 17-May   \\
\rowcolor{table-color-light}Estonia                & 28-Feb     & 07-May   & Mozambique & 23-Mar     & 12-Jul   & Ukraine             & 04-Mar     & 21-May   \\
\rowcolor{table-color-light}Ethiopia               & 14-Mar     & 10-Sep   & Mauritius  & 20-Mar     & 14-May   & USA                 & 21-Jan     & 14-Jun   \\
\rowcolor{table-color-light}Finland                & 30-Jan     & 13-Apr   & Malawi     & 03-Apr     & 31-Aug   & Venezuela           & 15-Mar     & 01-Nov   \\
\rowcolor{table-color-light}France                 & 25-Jan     & 25-May   & Malaysia   & 25-Jan     & 09-Jun   & Viet Nam            & 24-Jan     & 14-Apr   \\
\rowcolor{table-color-light}Gabon                  & 13-Mar     & 15-Oct   & Namibia    & 15-Mar     & 04-May   & South Africa        & 06-Mar     & 31-May   \\
\rowcolor{table-color-light}UK                     & 01-Feb     & 03-Nov   & Niger      & 21-Mar     & 12-May   & Zambia              & 19-Mar     & 07-May  \\ \hline
\end{tabular}%
}
\end{table}

Altogether, we end up with data on 28 variables and 105 countries. Table~\ref{tab:summary} reports the summary statistics of each variable. We hereby point out that as a measure of the correlate the log of the last observed value is taken (the value in 2019), unless otherwise stated in Table~\ref{tab:summary}. This prevents the possible problem of endogenous independent variables in the specification of the regression.

\begin{figure*}[t!]
\includegraphics[width=15cm]{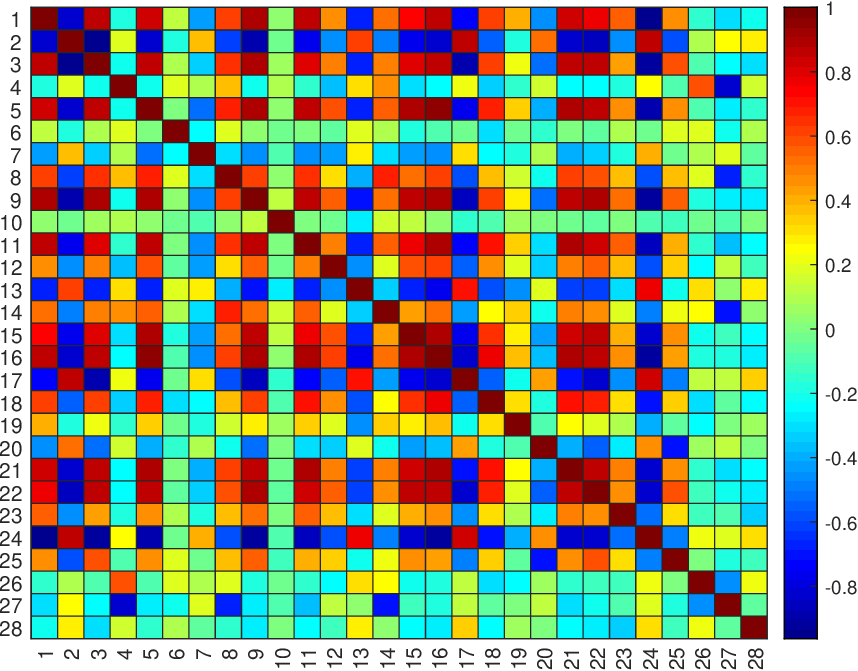}
\resizebox{\textwidth}{!}{
\begin{tabular}{|d{0}|l|d{0}|l|d{0}|l|d{0}|l|d{0}|l|d{0}|l|}\hline
\multicolumn{12}{|c|}{\textbf{Variable labels}}\\\hline
\rowcolor{table-color-light}  1.& Life expectancy           &  6.& Population density & 11.& Health coverage &  16.& Gov.h.spending &  21.& Digitalization & 26.& Comorbidities \\
\rowcolor{table-color-light}  2.&  Young population &  7.& Rural population &  12.& Migration &  17.& Household size &  22.&Medical resources& 27.  & Soc. Connectedness\\
\rowcolor{table-color-light}  3.&  Elderly population &  8.& Int. tourism  &  13.& Air pollution  &  18.& Overweight prevalence &  23.& Immunization&   28.& Gender\\
\rowcolor{table-color-light}  4.&  Population        & 9.& Education & 14.& Air transport &  19.& Asthma prevalence &  24.& Mortality& & \\
\rowcolor{table-color-light}  5.&  Econ. development   & 10.& Labor market  & 15.& Sus. development &  20.& Income inequality &  25.& Weather&  & \\   
\rowcolor{table-color-light}  &  & & & & &  & &  &  &  &
   \\\hline 
\end{tabular}}
\caption{\textbf{ Correlation matrix.
\label{fig:corr-matrix}}}
\end{figure*}

In Fig.~\ref{fig:corr-matrix} we plot the correlation matrix between the potential correlates. It can be observed, that in general the correlation between the variables is large. Out of 378 variable pairs, 102 have correlation that is either below -0.6 or above 0.6.

\begin{table}[t]
\caption{\textbf{Summary statistics.} \\  $^{a}$ Raw values. \\ $^{b}$ Individual calculations. \\ $^{c}$ 10 year averages. \label{tab:summary}}
\resizebox{0.90\textwidth}{!}{%
\begin{tabular}{|l|l|d{2}|d{2}|}
\hline
\multicolumn{1}{|c|}{\textbf{Variable}} & \multicolumn{1}{|c|}{\textbf{Measure}} & \multicolumn{1}{|c|}{\textbf{Mean}} & \multicolumn{1}{|c|}{\textbf{Std.}} \\\hline\hline
\rowcolor{table-color-light} Coronavirus outcome & Coronavirus infections p.m.p. & 5.92 &2.20 \\
\rowcolor{table-color-light} & Coronavirus deaths p.m.p. & 2.58 & 2.20 \\
\rowcolor{table-color-light}Government response & Government response index & -1.84 & 0.37\\
\rowcolor{table-color-light}Epidemic duration & Days since first registered local case$^{a}$ &99.97 & 61.43 \\
\rowcolor{table-color-light}&Days since first global case$^{a}$ & 58.22 & 19.20 \\
\hline
\rowcolor{table-color-dark}\multicolumn{4}{|c|}{\textbf{Healthcare Infrastructure}} \\\hline
\rowcolor{table-color-light} Medical resources & Medical resources index$^{b}$ & 0.10&1.06 \\
\rowcolor{table-color-light}Health coverage&UHC service coverage index&4.18&0.24 \\\hline
\rowcolor{table-color-dark}\multicolumn{4}{|c|}{\textbf{National health statistics}} \\\hline
\rowcolor{table-color-light}Life expectancy&Life expectancy at birth, (years)& 4.29&0.10 \\
\rowcolor{table-color-light}Mortality & Non-natural causes mortality index$^{b}$ & -0.30&1.04\\
\rowcolor{table-color-light}Immunization & Immunization index$^{b}$ &0.15&0.66 \\
\rowcolor{table-color-light} Comorbidities & Comorbidities index$^{b}$ &0.35&3.68\\
\rowcolor{table-color-light}Overweight prevalence& \% of adults with BMI $>$ 25 $kg/m^2$& 3.81 & 0.38 \\
\rowcolor{table-color-light}Asthma prevalence & Asthma prevalence (\% of population)& 1.53 & 0.33 \\\hline
\rowcolor{table-color-dark}\multicolumn{4}{|c|}{\textbf{Economic performance}} \\\hline
\rowcolor{table-color-light}Economic development&GDP p.c., PPP \$ &9.58&1.08 \\
\rowcolor{table-color-light}Labor market&Employment to population ratio (\%) &4.02&0.20 \\
\rowcolor{table-color-light}Government spending&Gov. health spending p.c., PPP \$$^{c}$& 5.85&1.71\\
\rowcolor{table-color-light}Income inequality& GINI index&3.61&0.21 \\
\hline
\rowcolor{table-color-dark}\multicolumn{4}{|c|}{\textbf{Societal characteristics}} \\\hline
\rowcolor{table-color-light}Social connectedness &Social connectedness index (PageRank)$^{b}$ & -0.79&1.31\\
\rowcolor{table-color-light}Digitalization &  Digitalization index$^{b}$ &0.11&0.95 \\
\rowcolor{table-color-light}Education& Human capital index&-0.55&0.27\\
\rowcolor{table-color-light}Household size& Avg. no. of persons in a household& 1.25&0.32 \\
\hline
\rowcolor{table-color-dark}\multicolumn{4}{|c|}{\textbf{Demographic structure}} \\\hline
\rowcolor{table-color-light}Elderly population&Population age 65+ (\% of total)&2.12&0.74 \\
\rowcolor{table-color-light}Young population&Population ages 0-14 (\% of total)&3.17&0.40 \\
\rowcolor{table-color-light}Gender&50\%+ male population&/&/ \\
\rowcolor{table-color-light}Rural population&Rural population (\% of total)&3.38&1.21 \\
\rowcolor{table-color-light}Migration&Int. migrant stock (\% of population)&1.09&1.47 \\
\rowcolor{table-color-light}Population density& People per sq. km &4.34&1.32 \\
\rowcolor{table-color-light}Population size&Population, total& 16.69&1.41 \\ \hline
\rowcolor{table-color-dark}\multicolumn{4}{|c|}{\textbf{Natural environment}} \\\hline
\rowcolor{table-color-light}Sustainable development& Ecological Footprint (gha/person)&0.98&0.65  \\
\rowcolor{table-color-light}Air Pollution&Yearly avg P.M. 2.5 &3.01&0.63 \\
\rowcolor{table-color-light}Air transport&Yearly passengers carried&8.22&2.74 \\
\rowcolor{table-color-light}International Tourism&Number of tourist arrivals &15.20&1.54 \\
\rowcolor{table-color-light}Weather (latitude)&Geographic coordinate: Latitude & 21.65 & 26.85 \\\hline
\end{tabular}}
\end{table}

\subsubsection{Individual indices}

\paragraph{Medical resources index:} The Medical resources index is estimated as a Principal Component Analysis (PCA) weighted index of the logs of three variables~\cite{vyas2006constructing}. These are:
\begin{itemize}
   \item[-] Physicians (per 1,000 people)
     \item[-] Nurses and midwives (per 1,000 people).
     \item[-] Hospital beds (per 1,000 people).
\end{itemize}

\paragraph{Non-natural causes mortality index:} The Non-natural causes mortality index is calculated as a Principal Component Analysis (PCA) weighted index of the logs of these four variables found in WDI:
\small{
\begin{itemize}
   \item[-] Mortality rate attributed to household and ambient air pollution, age-standardized (per 100,000 population).
     \item[-] Cause of death, by communicable diseases and maternal, prenatal and nutrition conditions (\% of total).
     \item[-] Mortality from CVD, cancer, diabetes or CRD between exact ages 30 and 70, female (\%).
     \item[-] Mortality rate attributed to unsafe water, unsafe sanitation and lack of hygiene (per 100,000 population).
\end{itemize}}

\paragraph{Immunization index:} The Immunization index is estimated as a Principal Component Analysis (PCA) weighted index of the logs of two variables:
\begin{itemize}
   \item[-] Immunization, DPT (\% of children ages 12-23 months).
     \item[-] Immunization, measles (\% of children ages 12-23 months).
\end{itemize}

\paragraph{Comorbidities index:} The Comorbidities index is calculated as a Principal Component Analysis (PCA) weighted index of the thirtheen individual measures describing the burdens of disease, measured by a metric called ‘Disability Adjusted Life Years‘ (DALYs). These are:
\begin{itemize}
   \item[-] Neglected tropical diseases and malaria.
     \item[-] Maternal disorders.
     \item[-] Neonatal disorders.
     \item[-] Nutritional deficiencies.
       \item[-] Neoplasms.
    \item[-] Cardiovascular diseases.
   \item[-] Chronic respiratory diseases.
     \item[-] Cirrhosis and other chronic liver diseases.
     \item[-] Digestive diseases.
     \item[-] Neurological disorders.
       \item[-] Mental and substance use disorders.
    \item[-] Musculoskeletal disorders.
\item[-] Other non-communicable diseases.
\end{itemize}

\paragraph{Social connectedness index:} The original social connectedness index (SCI) was introduced in~\cite{bailey2018social} as a measure of the magnitude of Facebook connections between pairs of countries $i$ and $j$. Formally, the $ij$-th index is estimated as
\small{
\begin{align}
   \text{Social Connectedness}_{ij} &= \frac{\text{FB Connections}_{ij}}{\text{FB Users}_{i} \times \text{FB Users}_{j}},
\end{align}}
where $\text{FB Connections}_{ij}$ is the total number of Facebook connections between $i$ and $j$ and $\text{FB Users}_{l}$ is the number of Facebook users in country $l$. Combining all pairs, this results in an $N \times N$ dimensional matrix. We transform it to be an only one-country measure by estimating the log of the PageRank (eigenvector centrality) of each country in the original SCI matrix~\cite{bonacich2007some}.

\paragraph{Digitalization index:} The Immunization index is estimated as a Principal Component Analysis (PCA) weighted index of the logs of four variables:
\begin{itemize}
   \item[-] Individuals using the Internet (\% of population).
     \item[-] Fixed broadband subscriptions (per 100 people).
     \item[-] Fixed telephone subscriptions (per 100 people).
     \item[-] Mobile cellular subscriptions (per 100 people).
\end{itemize}

\subsection{Bayesian model averaging}
\label{sec:BMA}

We specify our linear regression model $M_m$ for the final number of COVID-19 infections/death rates of the first wave as
\begin{align}
y_i &= \beta_0 +  \beta_m^T \mathbf{X}^m_i + \gamma s_i + \delta d_i + u_i, 
\label{eq:master}
\end{align}
where, for simplicity, we denote both the log of registered COVID-19 infections per million population and the log of COVID-19 deaths per million population of country $i$ as $y_i$. In the equation, $\mathbf{X}^m_i$ it is a $k_m$ dimensional vector of health, social and economic explanatory variables that determine the dependent variable, $\beta_m$ is the vector describing their marginal contributions, $\beta_0$ is the intercept of the regression, and $u_i$ is the error term. The $s_i$ term controls for the impact of government responses, and $\gamma$ is its coefficient. Finally, we also include the term $d_i$, with $\delta$ capturing its marginal effect, that measures the duration of the pandemics within the economy. This allows us to control for the possibility that the countries are in a different state of the disease spreading process.

BMA leverages Bayesian statistics to account for model uncertainty by estimating each possible
model $M_m$, and thus evaluating the posterior distribution of each parameter value and probability that a particular model is the correct one~\cite{moral2015model}. More precisely, in BMA, the posterior probability for the parameters $g(\beta_m|y, M_m)$ is calculated using $M_m$ as:
\begin{align}
    g(\beta_m|y, M_m) &= \frac{f(y|\beta_m, M_m) g(\beta_m|M_m)}{f(y|M_m)}.
    \label{eq:posterior-probability}
\end{align}
It is clear that the posterior probability is proportional to $f(y|\beta_m, M_m)$, -
the likelihood of seeing the data under model $M_m$ with parameters $\beta_m$, and $g(\beta_m|M_m)$ – the prior distribution of the parameters included in the proposed model. By assuming a prior model probability $P(M_m)$, we can implement the same rule to evaluate the posterior probability that model $M_m$ is the true one, as 
\begin{align}
    P(M_m|y) &= \frac{f(y|M_m) P(M_m)}{f(y)} = \frac{f(y|M_m) P(M_m)}{\sum_{n=1}^{2^k}f(y|M_n) P(M_n)}.
    \label{eq:model-probability}
\end{align}

The term $f(y|M_m)$ is called the marginal likelihood of the model and is used to compare different
models to each other. The posterior model probability can also be written as
\begin{align}
    P(M_m|y) &= \frac{B_{m0} P(M_m) }{\sum_{n=1}^{2^k} B_{n0} P(M_n)},
\end{align}
where $B_{m0}$ is the Bayes information criterion between model $M_m$ and the baseline model $M_0$. In our case this is the model including government social distancing measures and the length of the coronavirus crisis in the country.

With this setup, we can define the posterior distribution of $\beta$ as a weighted average of the posterior
distributions of the parameters under each model using the posterior model probabilities as weights
\begin{align}
    g(\beta | y) &= \sum_{j=1}^{2^k} g(\beta|y,M_m) P(M_m|y).
    \label{eq:beta-posterior}
\end{align}

Here, we are interested only in some parameters of the posterior distribution, such as the posterior
mean and variance of each parameter. Using equation~(\ref{eq:beta-posterior}) we can calculate the posterior mean as:
\begin{align}
    \EX \left[(\beta |y \right] &= \sum_{m=1}^{2^k}\EX \left[(\beta |y, M_m \right] P(M_m|y),
\end{align}
and the posterior variance as:
\begin{align}
    \mathrm{var} \left[(\beta |y \right] &= \sum_{m=1}^{2^k}   \mathrm{var} \left[(\beta |y, M_m \right] P(M_m|y) + \sum_{m=1}^{2^k} P(M_m|y) \bigg(\EX \left[(\beta |y, M_m \right] - \EX \left[(\beta |y, \right] \bigg)^2 .
\end{align}

Since the posterior mean is a point estimate of the average marginal contribution, we use it as our measure of the effect of the correlate on the COVID-19 impact.

Another interesting statistic is the posterior inclusion probability $PIP_h$ of a variable $h$, which
measures the posterior probability that the variable is included in the ‘true’ model. Mathematically, $PIP_h$ is defined as the sum of the posterior model probabilities for all of the models that include the variable:
\begin{align}
    PIP_h &= (P(\beta_h \neq 0) = \sum_{m :\beta_h \neq 0}^{2^k}P(M_m|y).
\end{align}

Posterior inclusion probabilities offer a more robust way of determining the effect of a variable in a model, as opposed to using p-values for determining statistical significance of a model coefficient because they incorporate the uncertainty of model selection. 

According to equations~(\ref{eq:posterior-probability}) and~(\ref{eq:model-probability}), it
is clear that we need to specify priors for the parameters of each model and for the model probability
itself. To keep the model simple and easily implemented here we use the most often implemented
priors. In other words, for the parameter space we elicit a prior on the error variance that is proportional
to its inverse, $p(\sigma^2) \approx 1/\sigma^2$, and a uniform distribution on the intercept, $p(\alpha) \to 1$, while the Zellner’s g-prior is used for the $\beta_m$ parameters, and for the model space we utilise the Beta-Binomial prior. To estimate the posterior parameters we use a Markov Chain Monte Carlo (MCMC) sampler, and report results from a run with 200 million recorded drawings and after a burn-in of 100 million discarded drawings. Finally, before we perform the inference the data for each variable is transformed into its z-score, in order to normalize the measuring unit. The theoretical background behind our setup can be read in Refs.~\cite{moral2015model,fernandez2001model,fernandez2001benchmark,ley2009effect}.

\subsection{Robustness checks}

\subsubsection{BMA outliers check}
\label{sec:BMA-robustness}
\begin{figure}
\begin{subfigure}{.5\textwidth}
  \centering
  \includegraphics[width=\linewidth]{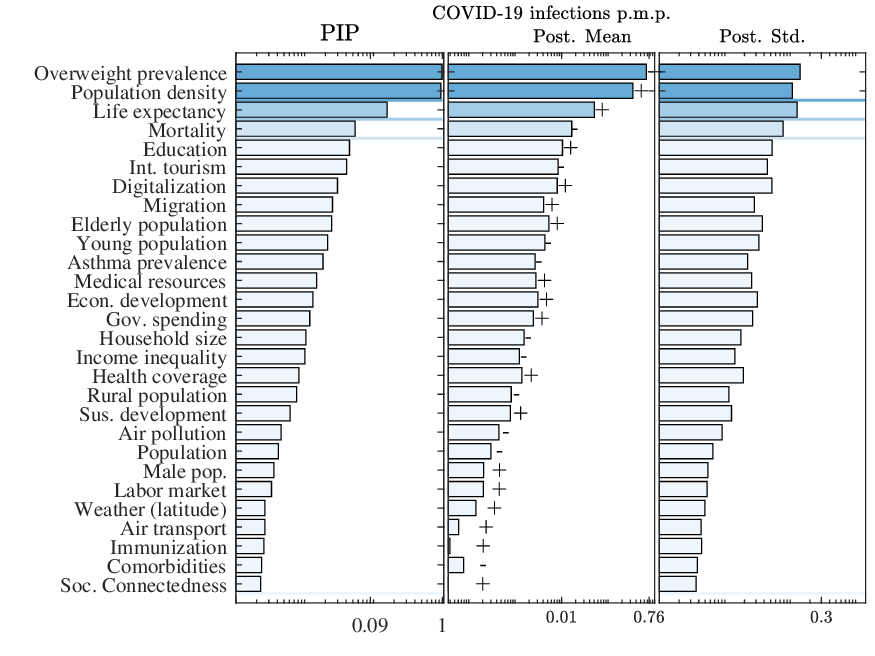}
\end{subfigure}%
\begin{subfigure}{.5\textwidth}
  \centering
  \includegraphics[width=\linewidth]{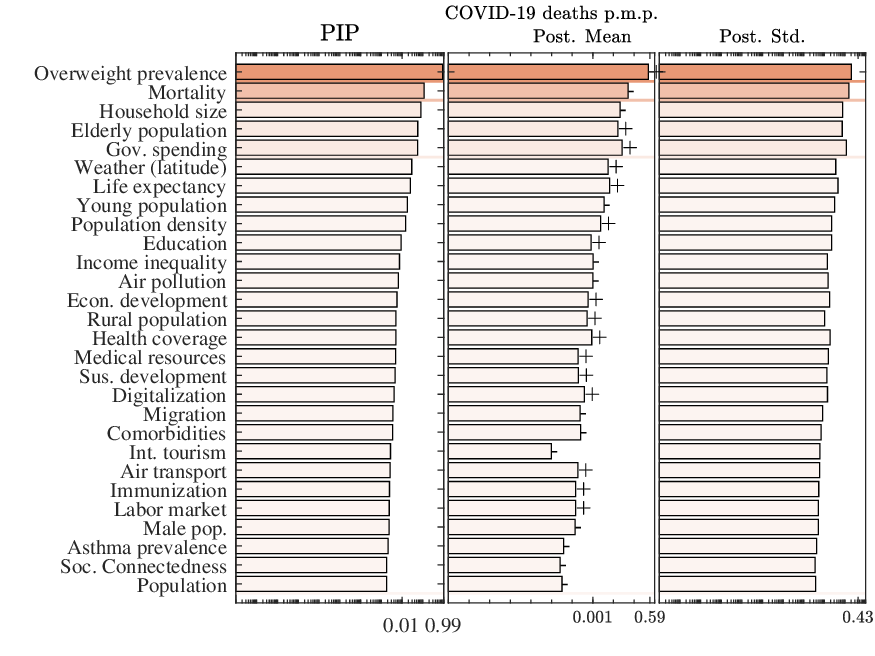}
\end{subfigure}
\caption{\textbf{BMA outliers check.}  Bars for the posterior inclusion probability (PIP), posterior mean (Post. Mean) and the posterior standard deviation (Post. Std.) of each potential correlate. The variables are ordered according to their PIP. The Post. Mean is in absolute value. The signs next to the bar of each variable indicate the direction of its impact. The horizontal lines divide the variables into groups according to their PIP. The horizontal axis is on a logarithmic scale. The setup for the estimation is described in SI Section~\ref{sec:BMA}.}
\label{fig:BMA-outliers}
\end{figure}

As said in the main text, we check the robustness of our results against the presence of outliers by removing a country from the sample and re-performing the BMA procedure with the resulting countries. We repeat this procedure for every country and recover the median results for each potential correlate. The results can be seen in Fig.~\ref{fig:BMA-outliers}. They are nearly identical to the ones presented in the main text, thus suggesting that our results are robust to outliers.

Table~\ref{tab:contributors} outlines the countries which had the biggest impact on the observed credibility of a given correlate. We define two types of countries, i) the weakest contributor, this is the country which when excluded from the sample leads to the largest PIP for the studied correlate; and ii) the strongest contributor -- i.e., the country which when excluded we observe the lowest PIP for the studied correlate. We find numerous countries which can be significant contributors for each correlate, thus indicating that there is indeed heterogeneity in the socio-economic features of the countries.

\begin{table}[H]
\caption{\textbf{Contributors to the credibility of a correlate.}  \label{tab:contributors}}
\resizebox{0.95\textwidth}{!}{%
\begin{tabular}{|l||c|c|c|c|}
\hline
& \multicolumn{2}{|c|}{\textbf{COVID-19 infections p.m.p.}} & \multicolumn{2}{|c|}{\textbf{COVID-19 deaths p.m.p.}} \\
\hline
\multicolumn{1}{|c||}{\textbf{Variable}} & \begin{tabular}{@{}c@{}}\textbf{Weakest} \\ \textbf{contributor} \end{tabular} & \begin{tabular}{@{}c@{}}\textbf{Strongest} \\ \textbf{contributor} \end{tabular} & \begin{tabular}{@{}c@{}}\textbf{Weakest} \\ \textbf{contributor} \end{tabular} &\begin{tabular}{@{}c@{}}\textbf{Strongest} \\ \textbf{contributor} \end{tabular} \\\hline\hline
\rowcolor{table-color-dark}\multicolumn{5}{|c|}{\textbf{Healthcare Infrastructure}} \\\hline
\rowcolor{table-color-light} Medical resources & Trinidad \& Tobago& Bolivia & Peru & Ireland \\
\rowcolor{table-color-light}Health coverage& Peru & Sweden& Peru & Sweden \\\hline
\rowcolor{table-color-dark}\multicolumn{5}{|c|}{\textbf{National health statistics}} \\\hline
\rowcolor{table-color-light}Life expectancy&Namibia&USA&Peru&Sweden \\
\rowcolor{table-color-light}Mortality & Peru&Jamaica&Peru&Jamaica\\
\rowcolor{table-color-light}Immunization &Jamaica&Peru&Sweden&Norway \\
\rowcolor{table-color-light}Overweight prevalence &USA&Peru&Jamaica&Peru\\
\rowcolor{table-color-light}Asthma prevalence &Peru&Bolivia&Sweden&Peru\\\hline
\rowcolor{table-color-dark}\multicolumn{5}{|c|}{\textbf{Economic performance}} \\\hline
\rowcolor{table-color-light}Economic development&Rwanda&Bolivia&Peru&Russia \\
\rowcolor{table-color-light}Labor market&USA&Trinidad \& Tobago&Russia&Mozambique \\
\rowcolor{table-color-light}Government spending&Rwanda&Jamaica&Peru&Jamaica\\
\rowcolor{table-color-light}Income inequality&USA&Brazil&Jamaica&Brazil \\
\hline
\rowcolor{table-color-dark}\multicolumn{5}{|c|}{\textbf{Societal characteristics}} \\\hline
\rowcolor{table-color-light}Social connectedness &USA&Italy&Sweden&Peru\\
\rowcolor{table-color-light}Digitalization &Peru&USA&Jamaica&Mozambique\\
\rowcolor{table-color-light}Education& Rwanda&Ireland&Peru&Australia\\
\rowcolor{table-color-light}Household size& Italy&Ireland&Peru&Ireland \\
\hline
\rowcolor{table-color-dark}\multicolumn{5}{|c|}{\textbf{Demographic structure}} \\\hline
\rowcolor{table-color-light}Elderly population&Italy&Rwanda&Italy&Sweden \\
\rowcolor{table-color-light}Young population&Italy&Sweden&Italy&Sweden \\
\rowcolor{table-color-light}Gender&USA&Peru&Mauritius&Slovakia \\
\rowcolor{table-color-light}Rural population&Rwanda&Italy&Russia&Sweden \\
\rowcolor{table-color-light}Migration&USA&Pakistan&Italy&Jamaica \\
\rowcolor{table-color-light}Population density&USA&Italy&Australia&Russia \\\hline
\rowcolor{table-color-dark}\multicolumn{5}{|c|}{\textbf{Natural environment}} \\\hline
\rowcolor{table-color-light}Sustainable development&USA&Togo&Italy&Mauritius \\
\rowcolor{table-color-light}Air Pollution&USA&Peru&Indonesia&Peru \\
\rowcolor{table-color-light}Air transport&Jamaica&Peru&Sweden&Mozambique \\
\rowcolor{table-color-light}International Tourism&Moldova&Peru&Bolivia&Ethiopia \\
\rowcolor{table-color-light}Weather (latitude)&Jamaica&Rwanda&Australia&Chile\\\hline
\end{tabular}}
\end{table}

\newpage
\subsubsection{Alternate end date of the first wave}
\label{sec:BMA-robustness2}
\begin{figure}[h!]
\begin{subfigure}{.5\textwidth}
  \centering
  \includegraphics[width=\linewidth]{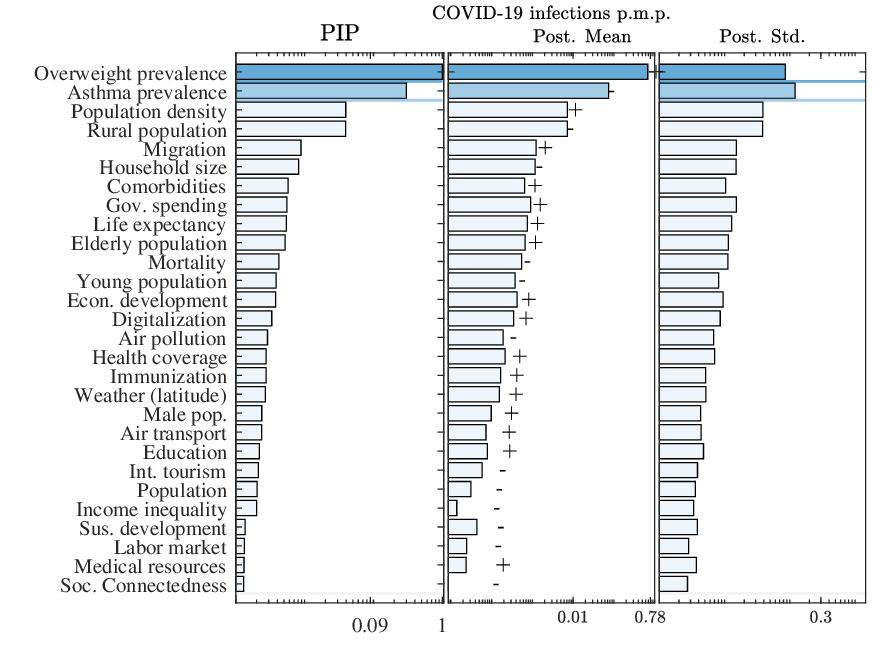}
\end{subfigure}%
\begin{subfigure}{.5\textwidth}
  \centering
  \includegraphics[width=\linewidth]{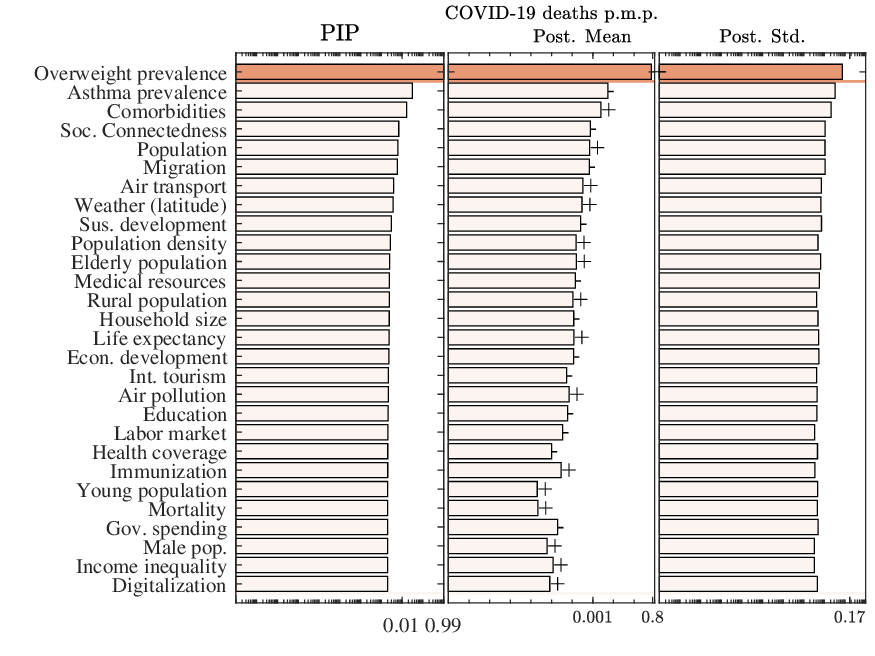}
\end{subfigure}
\caption{\textbf{BMA robustness results. The end date is the first time the gov. response index is 20\% lower than its maximum.} Bars for the posterior inclusion probability (PIP), posterior mean (Post. Mean) and the posterior standard deviation (Post. Std.) of each potential correlate. The variables are ordered according to their PIP. The Post. Mean is in absolute value. The signs next to the bar of each variable indicate the direction of its impact. The horizontal lines divide the variables into groups according to their PIP. The horizontal axis is on a logarithmic scale. The setup for the estimation is described in SI Section~\ref{sec:BMA}.}
\label{fig:BMA-results_rob_alternate_date}
\end{figure}

In this robustness check, we change the end date of the pandemic to be equal to the first date after the day at which the daily government response index is at its maximum and that is at least 20\% lower than the daily maximum. This effectively prolongs the duration of the first wave. The results are shown in Fig.~\ref{fig:BMA-results_rob_alternate_date}. In this case, it appears that there are more variables that are either strong or medium correlates of the COVID-19 infections/death rates. Nonetheless, the variables which were found in the main results, persist in being correlates with strong or medium evidence.

\subsubsection{BMA quasi-Poisson specification}

In this check, we change the dependent variable to be the raw number of infections and deaths at the end of the first wave. That is, now the dependent variable describes counts and the linear regression framework is not a suitable model. Instead, for the estimation of the marginal impact we use a quasi-Poisson model. This is the most often used procedure when the dependent variable is given as a count that has a large variance~\cite{ver2007quasi}. Indeed, the number of COVID-19 infections or deaths has a variance larger than its mean. This is apparently due to the disparate effect the pandemic had throughout the world.

\begin{figure}[t!]
\begin{subfigure}{.5\textwidth}
  \centering
  \includegraphics[width=\linewidth]{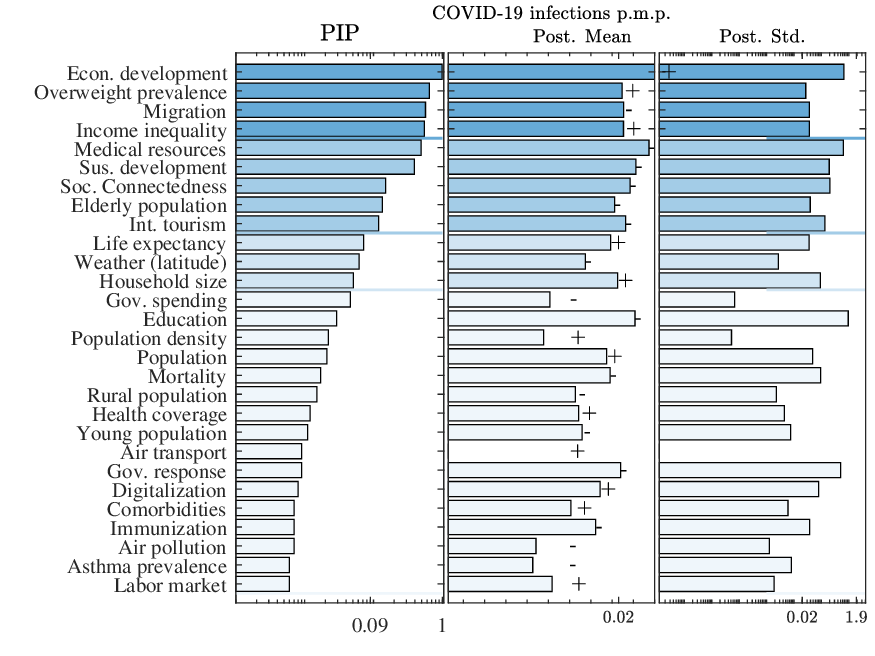}
\end{subfigure}%
\begin{subfigure}{.5\textwidth}
  \centering
  \includegraphics[width=\linewidth]{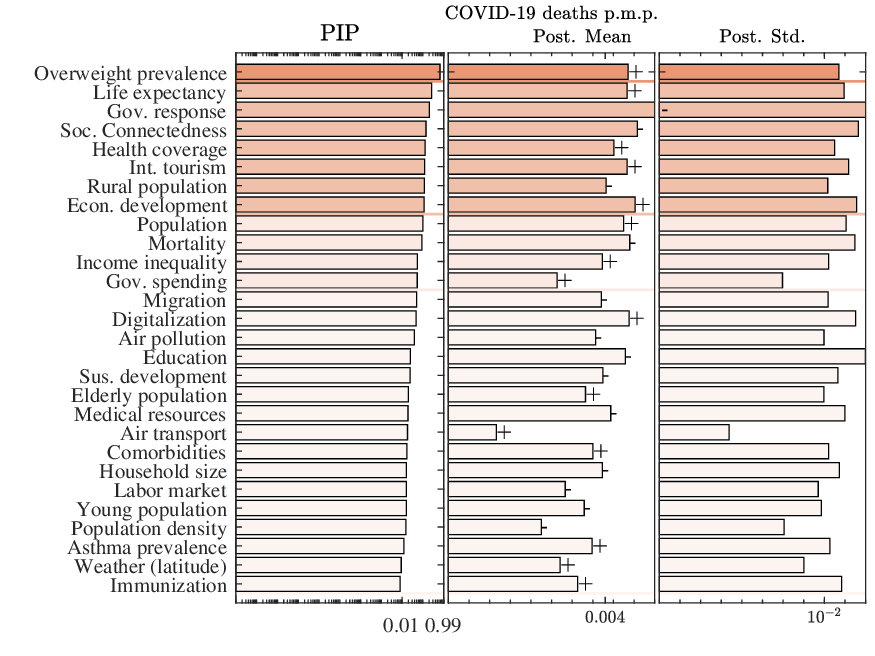}
\end{subfigure}
\caption{\textbf{BMA robustness results. The link function used in the model is quasi-Poisson.} Bars for the posterior inclusion probability (PIP), posterior mean (Post. Mean) and the posterior standard deviation (Post. Std.) of each potential correlate. The variables are ordered according to their PIP. The Post. Mean is in absolute value. The signs next to the bar of each variable indicate the direction of its impact. The horizontal lines divide the variables into groups according to their PIP. The horizontal axis is on a logarithmic scale. The setup used to estimate the results is described in SI Section~\ref{sec:BMA}.}
\label{fig:BMA-results_poisson}
\end{figure}

The results can be seen in Fig.~\ref{fig:BMA-results_poisson}. Again, the variables that were found to be strong correlates with the COVID-19 infections and mortality rates, remain strong correlates even in this specification. Thus, it can be concluded that our results are robust to a different model specification.

\subsubsection{BMA Spatial autocorrelation check}

In the last check we add a spatial weighting matrix in the baseline model in order to account for the potential spatial autocorrelation (SAR) in the spread of COVID-19. Multiple studies have indicated that this effect might exist (See for example~\cite{krisztin2020spatial}). Again our findings do not significantly change.

The SAR model which we use, takes the following matrix form
\begin{align}
    \mathbf{y} = \rho \mathbf{W} \mathbf{y} + \gamma \mathbf{s} +\mathbf{d} \delta   + \epsilon, 
\end{align}
Where $\mathbf{y}$ is the $n$-dimensional vector of infection or mortality rates, $\mathbf{s}$ and $\mathbf{d}$ are matrices containing the baseline independent variables and $\mathbf{W}$ is a known row-standardized spatial weight distance matrix between the studied countries. The parameter
$\rho$ is a coefficient on the spatially lagged dependent variable, $\mathbf{W} \mathbf{y}$. The spatial weight matrix, $\mathbf{W}$, is $n\times n$ stochastic matrix, where $n$ is the number of countries with element $w_{ij}$ defining the spatial relations between locations $i$ and $j$. This matrix is constructed through the following steps:

\begin{enumerate}
\item Gather data for the latitude and longitude of each county from \href{https://developers.google.com/public-data/docs/canonical/countries_csv}{Google Developers}.
\item Calculate the Haversine distance $D_{ij}$ between each pair of countries $i$ and $j$ using the data. This procedure allows us to determine the great-circle distance between any two countries on a sphere given their longitudes and latitudes  (See Ref.~\cite{robusto1957cosine}).
\item Construct a distance matrix, $\mathbf{D} = \left[ D_{ij}\right]$, between the countries using the estimations from the previous step.
\item Row-standardize the distance matrix to obtain the spatial weight matrix, $\mathbf{W}$. That is, the $ij$-th entry of the $\mathbf{W} = \left[ W_{ij}\right]$ is $W_{ij} = D_{ij}/ \sum_j D_{ij}$.
\end{enumerate}

The baseline SAR results are presented in Table \ref{tab:sar-baseline}. We observe that the spatial autocorrelation coefficient estimate for the SAR model is negative and statistically significant when the dependent variable is the infection rate, indicating the presence of spatial autocorrelation in the regression relationship. The coefficient remains negative when the dependent variable is the mortality rate, though it loses its significance.

\begin{table}[H]
\caption{\textbf{SAR results.} $*$ indicates significance at $\alpha = 0.05$ \label{tab:sar-baseline}}

\resizebox{0.80\textwidth}{!}{%
\begin{tabular}{|c|c|c|c|c|} \hline
                    & \multicolumn{2}{l|}{\textbf{COVID 19 infections p.m.p.}} & \multicolumn{2}{l|}{\textbf{COVID 19 deaths p.m.p.}}
 \\ \hline
        \textbf{Variable}                     & \textbf{Coefficient}           & \textbf{p-value}           & \textbf{Coefficient}            & \textbf{p-value}            \\ \hline
\rowcolor{table-color-light}Gov. response (log)          & -3.09                 &         0.00$*$          & -1.46                  &                   0.00*\\
\rowcolor{table-color-light}Days since first local case  & 0.02                  &          0.00$*$        & 0.02                   &                0.00*   \\
\rowcolor{table-color-light}Days since first global case & 0.05                  &          0.00$*$         & 0.01                   &                 0.51  \\
\rowcolor{table-color-light}Distance-based spatial weights          & -0.81                 &          0.03$*$         & -1.00                     &  0.13
\\ \hline
\end{tabular}}
\end{table}

\begin{figure}[t!]
\begin{subfigure}{.5\textwidth}
  \centering
  \includegraphics[width=\linewidth]{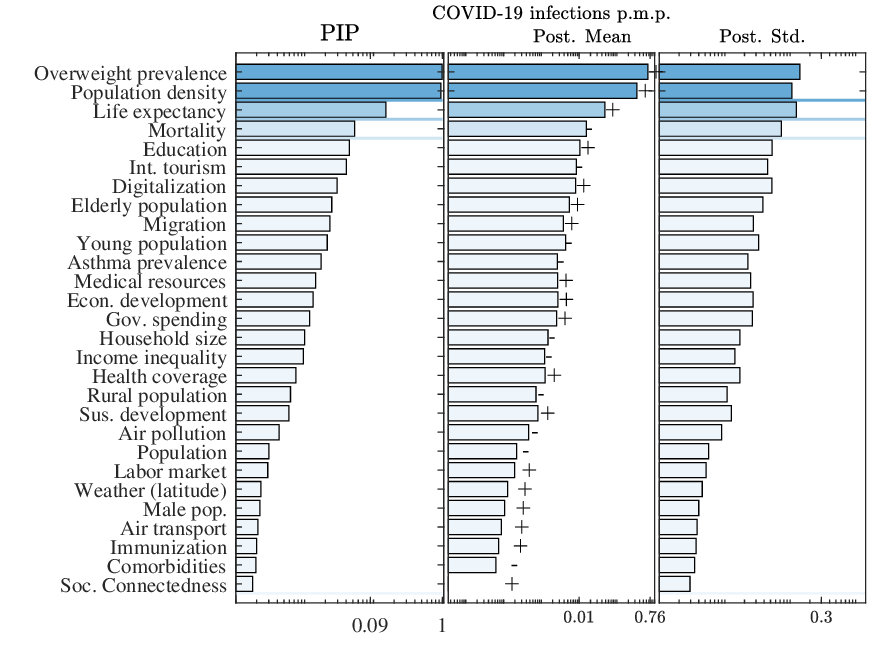}
\end{subfigure}%
\begin{subfigure}{.5\textwidth}
  \centering
  \includegraphics[width=\linewidth]{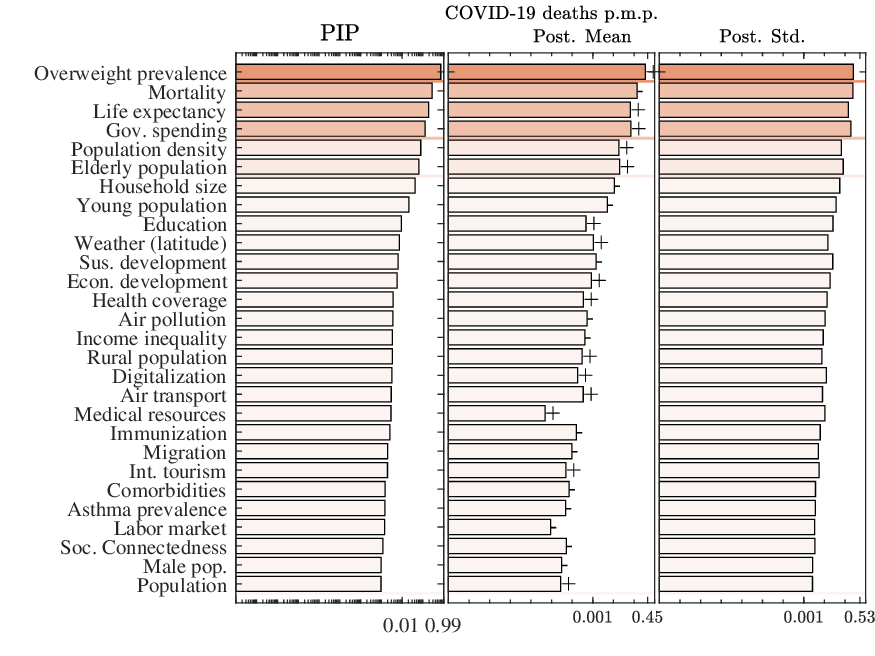}
\end{subfigure}
\caption{\textbf{BMA robustness results. The base model includes Spatial Autoregressive coefficient.}Bars for the posterior inclusion probability (PIP), posterior mean (Post. Mean) and the posterior standard deviation (Post. Std.) of each potential correlate. The variables are ordered according to their PIP. The Post. Mean is in absolute value. The signs next to the bar of each variable indicate the direction of its impact. The horizontal lines divide the variables into groups according to their PIP. The horizontal axis is on a logarithmic scale. The setup used to estimate the results is described in SI Section~\ref{sec:BMA}.}
\label{fig:BMA-results_spatial}
\end{figure}

The results for the BMA after implementing SAR as a baseline model, can be seen in Fig.~\ref{fig:BMA-results_spatial}. They are nearly identical to the ones presented in the main text, thus suggesting that our results are robust after accounting for spatial autocorrelation.

\newpage

\subsection{Construction of the coronavirus correlates Jointness space}
\label{sec:jointness}

To construct the coronavirus correlates Jointness space we utilize a network approach. In this network, the nodes represent the potential health, social and economic correlates, whereas the edge between a pair of correlates is given by a Jointness measure of the posterior probability that the pair is included in the same model explaining the COVID-19 infections/mortality rates. As a Jointness measure we utilize the the Hofmarcher et al. Jointness test. This test is a regularised version of
the well-known Yule’s Q association coefficient and is derived based on an augmented contingency table of variable inclusion. The table allows us to avoid the problems that arise
due to zero counts~\cite{hofmarcher2018bivariate}. The test statistic, $J_{hk}$ between variables $h$ and $k$, is calculated as
\begin{align}
    J_{hk} &= \frac{(a+\frac{1}{2}) (d+\frac{1}{2}) - (b+\frac{1}{2}) (c+\frac{1}{2})}{(a+\frac{1}{2}) (d+\frac{1}{2}) + (b+\frac{1}{2}) (c+\frac{1}{2})},
\end{align}
where $a$,$b$,$c$ and $d$ are the empirical counts of the MCMC drawings in which, respectively, $h$ and $k$ are included together; $h$ is included and $k$ is excluded; $h$ is not included and $k$ is included; and both $h$ and $k$ are excluded. The main advantage of this test over other jointness measures is that it is appropriately defined as long as one of the studied variables is included in the true model with positive probability. Moreover, it is monotonic, with larger values implying that the two variables are complements; commutative,i.e. $J_{hk} = J_{kh}$; it is bounded between $-1$, and $1$, and has an adequate limiting behavior.

To visualize the resulting network we use only the positive links (those that are greater than $0$). To set the coordinates of each node we use the Force-Layout drawing algorithm.

\end{document}